\documentclass[11pt,twocolumn,prl,aps,preprintnumbers,%amsmath,amssymb,
 superscriptaddress, showkeys,nofootinbib]{revtex4-1}
\usepackage{amsmath,amsfonts,amstext,txfonts,amssymb,
	hyperref,tcolorbox,color,ulem
	}

\usepackage[english]{babel}
\newcommand{\comillas}[1]{\textquotedblleft #1\textquotedblright}

\newcommand\redsout{\bgroup\markoverwith{\textcolor{red}{\rule[0.5ex]{2pt}{0.4pt}}}\ULon}
\begin{document}
\title{Mathematical model for the thermal enhancement of radiation response: Thermodynamic approach}

\author{Adriana M. De Mendoza*}
\affiliation{OncoRay - National Center for Radiation Research in Oncology, Faculty of Medicine and University Hospital C.G. Carus, TU Dresden, HZDR, Dresden, Germany}

\author{So\v{n}a Michl\'ikov\'a}
\affiliation{OncoRay - National Center for Radiation Research in Oncology, Faculty of Medicine and University Hospital C.G. Carus, TU Dresden, HZDR, Dresden, Germany}

\author{Johann Berger}
\affiliation{ICCAS - Innovation Center Computer Assisted Surgery, University of Leipzig, Leipzig, Germany}

\author{Jens Karschau}
\affiliation{OncoRay - National Center for Radiation Research in Oncology, Faculty of Medicine and University Hospital C.G. Carus, TU Dresden, HZDR, Dresden, Germany}

\author{Leoni A. Kunz-Schughart}\thanks{These authors contributed equally to this work}
\affiliation{OncoRay - National Center for Radiation Research in Oncology, Faculty of Medicine and University Hospital C.G. Carus, TU Dresden, HZDR, Dresden, Germany}
\affiliation{National Center for Tumor Diseases (NCT), partner site Dresden, Germany}

\author{Damian D. McLeod}\thanks{These authors contributed equally to this work}
\affiliation{OncoRay - National Center for Radiation Research in Oncology, Faculty of Medicine and University Hospital C.G. Carus, TU Dresden, HZDR, Dresden, Germany}
\affiliation{School of Biomedical Sciences and Pharmacy, Faculty of Health and Medicine, Hunter Medical Research Institute, The University of Newcastle, Callaghan, Australia}

%=======================================================================================================================================================================================
%=======================================================================================================================================================================================
\begin{abstract}{
%\vspace{1.0cm}
Radiotherapy can effectively kill malignant cells, but the doses required to cure cancer patients may inflict severe collateral damage to adjacent healthy tissues. Hyperthermia (HT) is a promising option to improve the outcome of radiation treatment (RT) and is increasingly applied in hospital. However, the synergistic effect of simultaneous thermoradiotherapy is not well understood yet, while its mathematical modelling is essential for treatment planning. To better understand this synergy, we propose a theoretical model in which the thermal enhancement ratio (TER) is explained by the fraction of cells being radiosensitised by the infliction of sublethal damage through mild HT. Further damage finally kills the cell or inhibits its proliferation in a non-reversible process. We suggest the TER to be proportional to the energy invested in the sensitisation, which is modelled as a simple rate process. Assuming protein denaturation as the main driver of HT-induced sublethal damage and considering the temperature dependence of the heat capacity of cellular proteins, the sensitisation rates were found to depend exponentially on temperature; in agreement with previous empirical observations. Our predictions well reproduce experimental data from in-vitro and in-vivo studies, explaining the thermal modulation of cellular radioresponse for simultaneous thermoradiotherapy. \\

\textit{*am.de259@uniandes.edu.co}

}
\end{abstract}

\maketitle

\section{Introduction}

Despite considerable efforts for decades towards the improvement of early diagnosis and therapy, cancer has remained a serious global health problem, with 18.1 million new cases and 9.6 million cancer deaths reported worldwide, just in 2018 \cite{stats}. Since the 1980s, mild hyperthermia (heating tumour tissue to 40.0 - 42.5 $^\circ$C for $\sim$ 1 h) is known to enhance the therapeutic outcomes in cancer patients, when combined with radio-, chemo- and/or immunotherapy \cite{Yagawa,Cheng}. Technological improvements in precise medical heating, imaging and non-invasive thermometry over the past decade have revived hyperthermia treatment (HT) as a precision cancer therapy \cite{Paulides,Cheng,Kang,Soares}, particularly when used in simultaneous combination with ionizing radiation \cite{Moros,Oberacker,Kosterev}. The number of ongoing HT clinical trials, either alone or in combination with different treatment modalities, evidences the increasing use of therapeutic HT (467 still ongoing clinical trials out of 1198 since 2000) \cite{CT}. Radiotherapy (RT) is supposedly a curative treatment modality, but the radiation dose required to eradicate all cancer cell subpopulations in a tumour can often not be applied due to severe acute or long-term side effects, which include radiation-induced tissue fibrosis and second malignancies \cite{Dracham}. Hyperthermia is known to be one of the most potent radiosensitisers \cite{Mei,Overgaard2013,Elming,Carbon,Neutrons}, meaning that less radiation is required to achieve the same local tumour cell kill, thereby reducing the adverse effects of radiation in the adjacent normal tissues e.g. \cite{vanderZee,Datta2016,Hurwitz,Datta20162,Bakker}.\\ 

The efficiency of combined HT+RT treatment clearly depends on the scheduled sequence of the two types of treatment and the time interval between them; best outcome for the patient is expected from a simultaneous application \cite{van_Leeuwen20172,van_Leeuwen2017,Overgaard2013,Overgaard}. However, several theoretical and practical problems still need to be overcome to implement simultaneous HT+RT approaches in routine clinical practice worldwide \cite{Spirou}. Indeed, in practical terms simultaneous treatment has remained challenging because spatially precise hyperthermia delivery is required to avoid unspecific synergistic, cytotoxic effects of HT+RT on the surrounding normal tissue, which would critically limit 

\begin{tcolorbox}[colback=gray!30]
\begin{center}
\textbf{Definitions box}\\
The key biological terms used in this work have been specified as follows:
\end{center}

\begin{small} 

$\blacklozenge$ \textbf{Cell kill (\comillas{dead state}):} From the radiobiological perspective, a cell is considered to be dead (killed) when it loses its proliferative capacity, i.e. is no longer able to divide (becomes replication-incompetent). This encompasses not only cells losing their membrane integrity and truly dying (by apoptosis, necrosis, or other), but also living cells undergoing terminal differentiation, permanent cell cycle arrest or senescence. This type of \textit{cell kill} leads to control of the malignant disease, independent of the underlying process.

\vspace{0.5cm}
$\blacklozenge$\textbf{Cell survival (\comillas{alive state}):} A cell is considered to survive if it remains replication-competent, i.e. retains its proliferative capacity after the treatment.

\vspace{0.5cm}
$\blacklozenge$\textbf{Cell damage:} Any type of deterioration of the cellular processes, regardless of origin, that advances the cell towards the \textit{dead state}.

\vspace{0.5cm}
$\blacklozenge$ \textbf{Radiological parameters $\alpha$ and $\beta$:} They characterize the radiosensitivity of cells or tumours. 

\vspace{0.2cm}
-$\alpha$ Characterizes the initial slope of logarithmic survival curves. It is associated to the mean number of DNA double strand breaks produced with a single radiation event \cite{Chadwick}.

\vspace{0.2cm}
-$\beta$ Characterizes the shoulder of logarithmic survival curves. It is associated to the mean number of DNA double strand breaks produced with two radiation events, i.e. two independent single strand breaks in close proximity that lead to formation of a double strand break \cite{Chadwick}.

\vspace{0.2cm}
-$\alpha/\beta$ \textit{ratio} Quantifies radiation sensitivity of tissue. The higher the ratio, the lower the sensitivity.

\vspace{0.5cm}
$\blacklozenge$ \textbf{Thermal enhancement ratio (TER):} Ratio between the radiation dose required to achieve a specific endpoint with ionizing radiation alone, and the radiation dose required to achieve the same endpoint in combination with hyperthermia.

\end{small} 
\end{tcolorbox}

\noindent the therapeutic benefit. Current standard clinical equipment is still not well-suited for such simultaneous and precise thermoradiotherapy. Accordingly there are only a few reports/completed trials, in which both forms of radiation were concomitantly applied in patients \cite{Peeken,Datta2015,Varma,Moros,Myerson1999}. From the theoretical perspective, mathematical models to predict the therapeutic outcome of various combinatorial treatment schemes are essential for a better understanding of the synergistic potential and therapeutic window of the two sources of energy (HT and RT), and are highly relevant to the design of adequate and individualized treatment planning in the clinical setting \cite{Datta,Crezee}. \\

Several mathematical models for individual RT and HT have been proposed, but there is poor consensus when it comes to the efficacy of combined treatment regimes. For RT, the LQ-model is the most extensively used approach to predict the effect of irradiation on cell populations \cite{reviewRT,RBbook}. This model describes the surviving fraction of cells as a function of the applied radiation dose $D_R$ by means of two main variables, called \comillas{radiological parameters} $\alpha$ and $\beta$ \cite{RBbook}. In the context of radiobiology, \comillas{survival} means the conservation of the cell's proliferative capacity \cite{Zaider} (see definitions box). Regarding HT, there is considerable literature describing the impact of heat on different cellular components \cite{Lepock1,Lepock2,Grimm,Kampinga3}, and several models are aimed to predict the survival of cells under HT treatments \cite{Pearce,Jung}. For thermal-radiosensitisation using temperatures of 40-46 $^\circ$C, there is a general agreement on a relevant role of DNA repair impairment by heat-induced protein denaturation in the processes of radiosensitisation between 40-46 $^\circ$C \cite{Mei,Kampinga3,Lepock1,reviewRT,Lepock2,Oei}. The majority of previous approaches to model the combined efficacy of hyperthermia and radiation on mammalian cells have implemented the thermal effects in the LQ-model by proposing empirical temperature dependencies for the radiological parameters \cite{Bruningk1,Bruningk2,vanLeeuwen1}, but the physical principles and the detailed mechanisms underlying this empirical dose-lowering concept are still elusive \cite{Oei}. The link between modelling concepts and plausible mechanistic explanations still needs to be established to serve as a more reliable framework for predictions. \\

Here, we propose a survival model for the simultaneous application of HT and RT that provides insights from a thermodynamic perspective. In our model, the modulation of the radiological parameters arises directly from the definition of the \textit{thermal enhancement ratio} (TER). It compares the radiation dose required to achieve a specific endpoint with ionizing radiation alone ($D_R$), e.g. surviving fraction of cells or tumour control probability, and the radiation dose required to achieve the same endpoint in combination with hyperthermia ($D_{R+H}$) $\text{TER}=\frac{D_R}{D_{R+H}}$ \cite{Gillette}. We propose the enhancement to be a rate limiting process, proportional to the energy invested in sensitising a cell to die. Our approach presents a theoretical basis to understand how hyperthermia results in radiosensitisation, a process that depends on treatment time and temperature. Our findings are consistent with previous experimental studies in the range of RT combined with mild hyperthermia.

\section{Results and discussion:}\label{Mresults}
\vspace{-0.5cm}
\begin{center}
\textbf{Mathematical model for the outcome of simultaneous HT+RT}
\end{center} 

In the following we describe our theoretical model and its correspondence with different types of experimental data in the range of mild HT derived from mammalian cell models. Numerous \textit{in-vitro} and \textit{in-vivo} studies reveal the successful and promising combinatorial application of HT and RT for anticancer treatment (see e.g. Refs \cite{Franken,Peeken,Datta2015}). However, the majority of the documented data is quite limited or incomplete and thus insufficient to test our model. To this end, we chose three rather dated seminal studies because, to our knowledge, they are the only ones which compile complete sets of thermal enhancement ratios, systematically obtained for several temperatures or treatment times in the HT regime. Two of these data sets were collected in course of \textit{in-vitro} 2D culture  experiments using  Chinese hamster ovary (CHO) \cite{Dikomey} and murine mammary carcinoma (M8013) cells \cite{Havemann}, respectively. Another set of data comes from an animal study with C3H murine mammary carcinoma xenografts \cite{Overgaard}.\\

%-----------------------------------------------------------------------------------------------------\\

\subsection{Hyperthermia affects the radiation dose-response curve} 

\begin{figure*}
	\centering
	\includegraphics[scale=0.4]{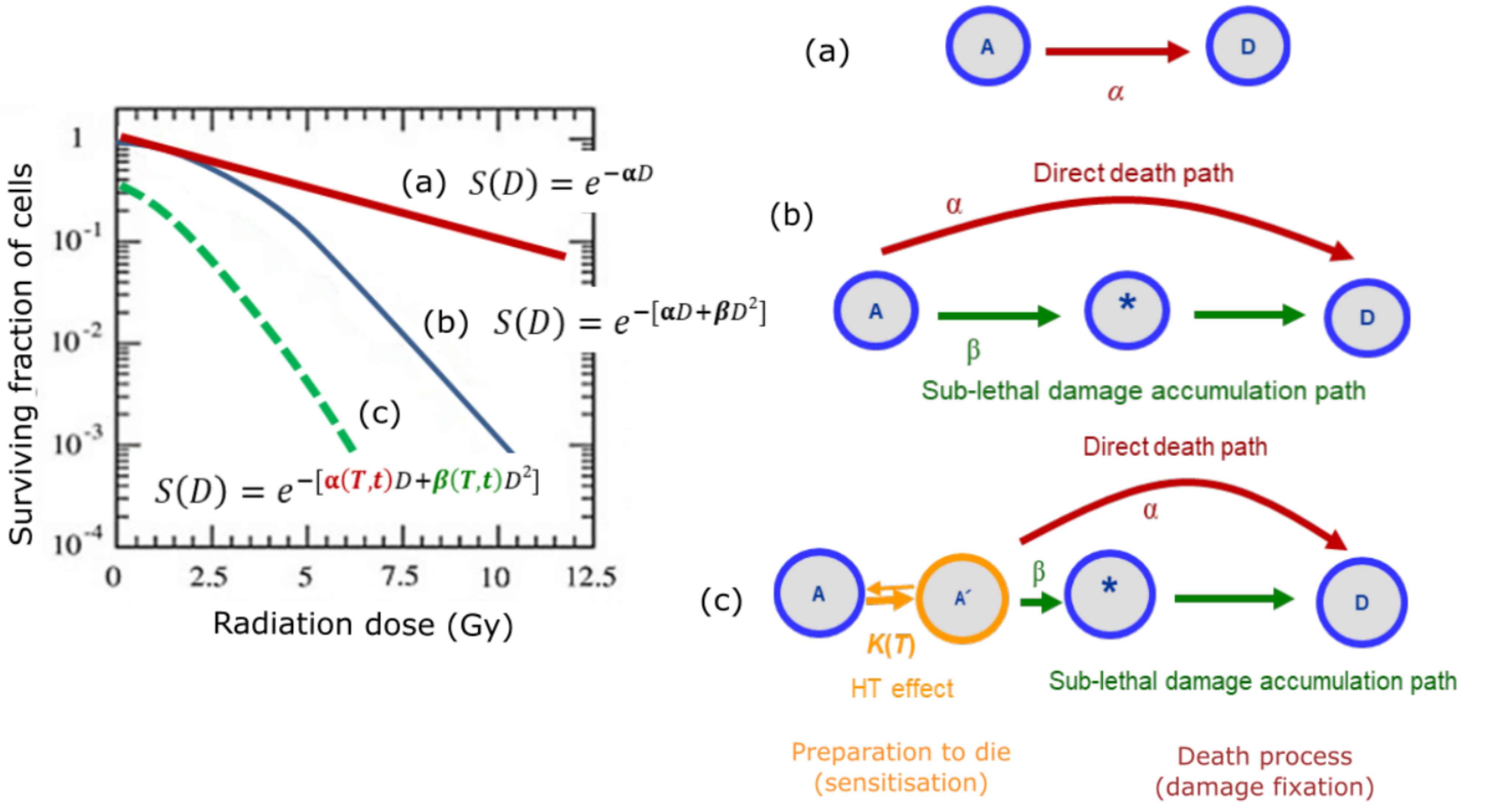}
	\caption{left: Schematic survival probabilities for the three cases depicted on the right. (a) Cell killing as a single rate process with transition rate from alive (A) to dead (D) $\alpha$. (b) Two-step cell killing process in the LQ-model for radiation. A cell transits from the alive state (A) to the dead state (D) through two possible paths: $\alpha$ for direct killing (a single hit suffices to kill), and $\beta$ for indirect killing (when two hits are required to kill). (c) Combined HT+RT: HT-induced damage elevates cells from state (A) to an activated state (A'), effectively reducing the $\alpha / \beta$ ratio. Since $\beta$ is more efficiently reduced, the direct path $\alpha$ dominates the killing process and consequently reduces the survival probability.}
	\label{fig1}
\end{figure*}

The LQ-model for radiotherapy predicts the surviving fraction of cells as an exponential function of the radiation dose, $S(D_R)=\exp\left\lbrace -(\alpha D_R + \beta D_R^2)\right\rbrace $ \cite{RBbook}. When HT is applied in combination with RT the parameters $\alpha$ and $\beta$ are modulated by both the temperature $T$, and the application time $t$ of heat \cite{Franken,Havemann,Kok,Myerson2004}. As a result, the sensitivity of cells to RT is increased and the radiation dose $D_{R+H}$ required to produce the same surviving fraction is lower. HT affects the survival probability curves in three ways: 1. the curves are shifted down as a consequence of cell killing from HT itself (offset at $D_R=0$), 2. there is a steeper initial slope ($\alpha$), and 3. the shoulder of the curve ($\beta$) is changed as illustrated in Fig.\ref{fig1}. In this work, the term \comillas{cell kill} is defined as the complete loss of proliferative capacity of a cell, regardless its membrane integrity.\\

%\onecolumngrid

%\twocolumngrid

The most accepted hypothesis for the radiosensitising effect of HT assumes the heat-induced denaturation of repair proteins impairs the DNA repair process upon irradiation \cite{Mei,reviewRT,Oei}. In the LQ-model hyperthermia mainly affects $\beta$, which is supposedly related to repairable DNA single-strand breaks (SSB), and the HT-induced sensitisation is generally associated with inhibition of DNA repair \cite{Datta}. Nevertheless, this description is incomplete because the change in $\alpha$ is not negligible. Given that $\beta$ is not exclusively related to pairs of SSB but also to clusters of DNA lesions \cite{RBbook}, we propose to differentiate between repairable and sublethal DNA damage, which are not necessarily the same. We suggest to extend the hypothesis of repair inhibition to a more general explanation based on sublethal damage accumulation (whether reversible or not), to better understand the synergy between radiation and thermal energy when applied to biological tissue. \\

\subsection{Modulation of $\alpha$ and $\beta$ by HT as a function of TER}

A portion of the thermal energy of HT goes into direct cell killing and another portion into radiosensitisation. For HT used for combinatorial therapy (40-46 $\circ$ C) only a minor fraction relates to direct cell-killing.\\

We therefore propose that the radiosensitising portion of the energy is invested in the accumulation of sublethal damage, facilitating radiation-induced cell death. In our model hyperthermia causes the cells to advance from an original undamaged state (A) to a more damaged state (A') in the sequence of sublethal damage (SLD) accumulation, as is illustrated in Fig. \ref{fig1}(c). Starting from (A') instead of (A), the radiation energy required to produce lethal and sublethal transitions is reduced, and hence, $\alpha$ and $\beta$ are effectively rescaled to $\alpha^*$ and $\beta^*$. Further, we assume that this modulation comes directly from the definition of the TER, in such a way that the new parameter ($\alpha^*$ and $\beta^*$) become treatment-time and temperature dependent (see Methods section for details)

\begin{align}
\label{a} \alpha^*(T,t)&=\alpha* \text{TER}\\
\label{b} \beta^*(T,t)&=\beta* \text{TER}^2.
\end{align}

\subsection{Thermodynamic basis of TER}

TER is expected to be proportional to the thermal energy absorbed by the cell, which is invested in the transition from (A) to (A') (transition towards 'dead state'). We propose this energy to increase linearly with the time of heat exposure $t$, and with the rate of energy absorption $k_E(T)$. In a simplified version of the SLD accumulation induced by hyperthermia, the step from (A) to (A') is represented by a single rate process, with a net rate of transition $k(T)$ (proportional to the rate of energy absorption) as depicted in Fig. \ref{fig1}(c) and expressed by

\begin{equation}\label{TF}
\text{TER}= o +a\:tk(T).
\end{equation}

Here, $o$ is the onset of the thermal enhancement ratio, which should converge to one for no HT treatment ($t=0$); $a$ is a parameter that accounts for the tumour size (or the amount of malignant cells) and the intrinsic sensitivity of the cells to RT and HT, and $k(T)$ is the temperature-dependent rate of the sensitisation process. Based on the thermodynamics of protein denaturation, we found the transition rate of this process to increase exponentially with increasing temperature $k(T)=c\:e^{b(T-T_g)}$ (see methods section for details of the model). Such exponential behavior with $(T-T_g)$ has been observed in previous works, but could not be explained \cite{Bruningk1,Overgaard6,Havemann}. In this equation, $c$ and $b$ are cell-type dependent parameters and $T_g$ is the dominant transition temperature, i.e. the average melting point of cellular proteins undergoing denaturation. We achieve this theoretical prediction by considering the change of the heat capacity of the proteins as a linear function of the temperature, and not as a constant value as usually assumed in Arrhenius kinetics. The heat capacity of cellular proteins displays a Lorentzian-type function of the temperature \cite{Lepock1,Lepock2}, which can be approximated at first order as linear functions in the vicinity of the melting point \cite{Prabhu,Atkins}. Remarkably, the melting point $T_g$ in both cases has good correspondence to the calorimetry studies performed by Lepock and collaborators \cite{Lepock1,Lepock2} where they found the melting point in the mild-hyperthermia treatment to be in the range of $45-48^{\circ}$C for different mammalian cells. Plugging the obtained transition rate into Eq.\ref{TF}, the TER reads

\begin{equation}\label{TF2}
\text{TER}=o+a't\:e^{b(T-T_g)},
\end{equation}

\noindent with $a'=ac$ for simplicity. This model predicts exponential increase of $\alpha*$ and $\beta*$ with temperature, which is much more pronounced for $\beta*$. These predictions are consistent with experimental results in cell cultures \cite{Dikomey,Havemann}, and data from human clinical trials \cite{Datta,Franken}.\\

Radiosensitising effects are also reflected and quantified by reductions in the $\alpha/\beta$ ratio, which is basically higher for intrinsically more radioresistant cells \cite{RBbook,Datta}. For the combined RT+HT scheme the $\alpha/\beta$ ratio is reduced as a consequence of the enhancement of the sublethal damage over the direct damage. The ratio for the combined treatment then reads $\alpha^*/\beta^*=\frac{\alpha/\beta}{\text{TER}}$.\\

\subsection{Predictions of experimental data from literature}

We tested the performance of our model (Eq.\ref{TF2}) on three experimental data sets that document thermal enhancement values in simultaneous HT+RT treatments for different temperatures. In two of them TER was measured for different treatment times and temperatures, and the third data set presents $\alpha$ and $\beta$ values obtained for various temperatures but just one treatment time. \\

\begin{figure*}
	\centering
	\includegraphics[scale=0.35]{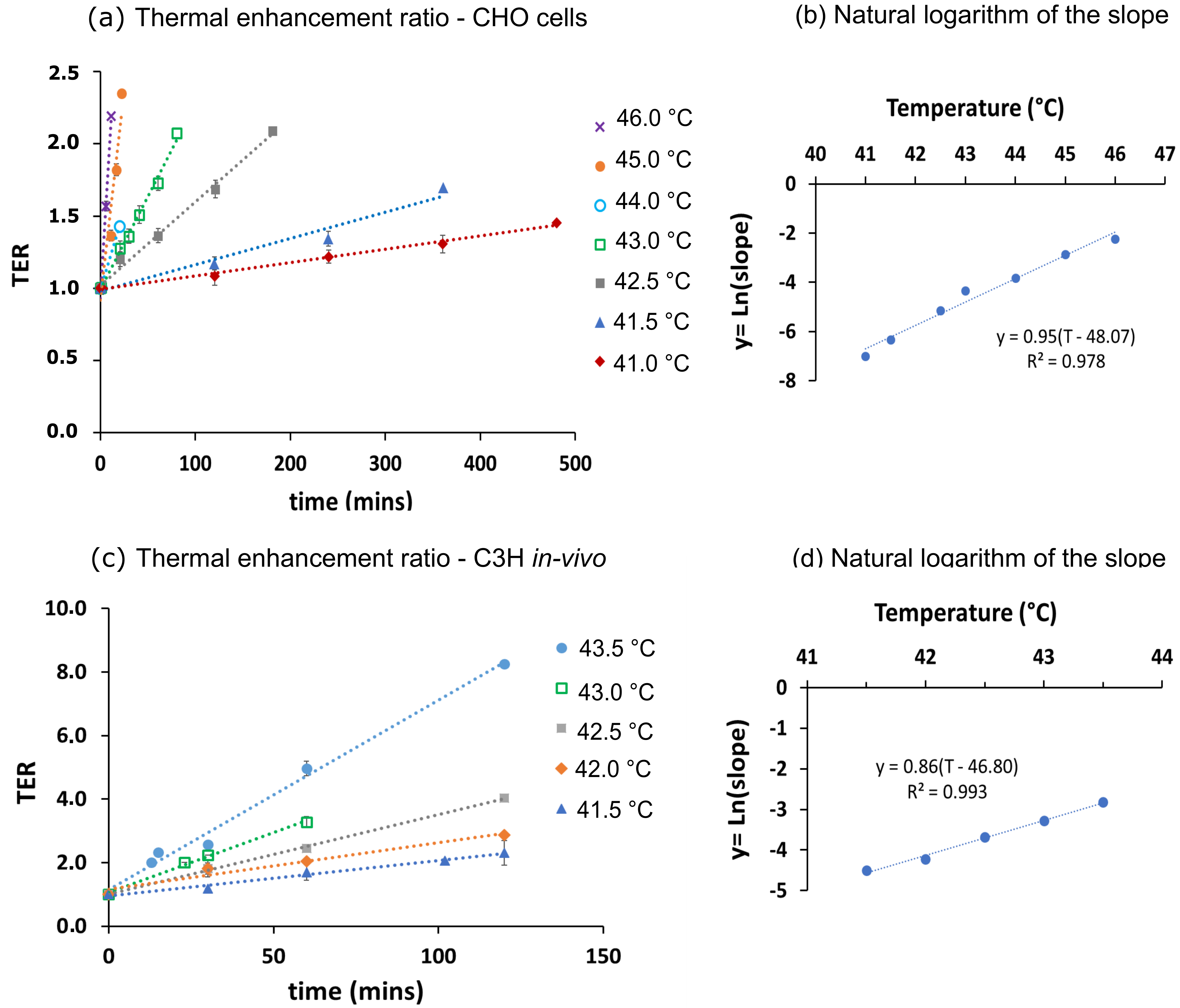}
	\caption{(a) and (c) show the linear dependency of the thermal enhancement ratio (TER) on time of exposure for CHO cells \textit{in vitro} and C3H mammary carcinoma cells in mice xenografts \textit{in vivo}, respectively. The slope of the linear fitting clearly depends on the temperature of the hyperthermia treatment, and the natural logarithm of the slope was plotted as a function of temperature for both datasets in (b) and (d). The linear trend lines show the exponential behaviour of the temperature dependent rate $k(T)$ according to Eq.\ref{TER}. The data for CHO cells (a-b) and C3H mammary carcinoma (c-d) was extracted from \cite{Dikomey} and \cite{Overgaard}, respectively.}
	\label{fig2}
\end{figure*}

\textit{Thermal enhancement ratio:} The first data set was recorded in \textit{in-vitro} 2D cell culture experiments (CHO cell line) \cite{Dikomey}, and the second one derived from an \textit{in-vivo} animal study (C3H mammary carcinoma tumour xenograft mouse model) \cite{Overgaard}. For both datasets we found that our model well predicts the outcome of these studies. As shown in Fig.\ref{fig2}(a) and Fig.\ref{fig2}(c), both datasets display a linear dependency of the TER with treatment time $t$ for all tested temperatures, indicating a rate-dependent behaviour of the function. The slope of each linear function is proportional to a temperature-dependent rate, matching the exponential fit shown in Fig.\ref{fig2}(b) and Fig.\ref{fig2}(d) respectively. The parameters and the respective coefficients of determination $R^2$ are summarized in Table \ref{tab1} for both examples. \\

%\onecolumngrid

%\twocolumngrid
\begin{table}[h]
\centering
%\begin{longtable}[H] {|p{2cm}|p{2.8cm}|p{2.8cm}|p{2.8cm}|}
\caption{Parameters of the TER model Eq.\ref{TF2}, obtained from CHO and C3H cell models \cite{Dikomey,Overgaard}}
\begin{tabular}{| c | c | c | c | c | c |}
\hline
\textbf{Cell model}     &   \textbf{o}  &   \textbf{$a'$}   &\textbf{$b$}   &   \textbf{$T_g$[$^\circ$C]}   &   \textbf{$R^2$} \\ \hline
CHO (\textit{in vitro}) & 0.97 $\pm 0.03$ & 1.00 & 0.95 & 48.07 & 0.978 \\
C3H (\textit{in vivo})  & 1.07 $\pm 0.04$ & 1.00 & 0.86 & 46.80 & 0.993 \\ \hline
\end{tabular}
\label{tab1}
\end{table}

\textit{Thermal modulation of $\alpha$ and $\beta$:} The third data set was documented in cultured M8013 murine mammary carcinoma cells \cite{Havemann}. In this study, the normal (non-thermotolerant) cell-line was compared with a thermotolerant modification. The authors determined the radiobiological parameters $\alpha$ and $\beta$ for cells irradiated halfway through a 30 min hyperthermia treatment (temperatures from 42 $^{\circ}$C to 47  $^{\circ}$C). In this case, we calculated the thermal enhancement of $\alpha$ and $\beta$ from Eqs. \ref{a} and \ref{b} to test our model:

\begin{equation*}
TER_{\alpha}=\frac{\alpha(T)}{\alpha} \hspace{0.5cm}\text{and} \\
\end{equation*}
\begin{equation}\label{terab}
TER_{\beta}=\sqrt{\frac{\beta(T)}{\beta}}.
\end{equation}

Here, $\alpha$ and $\beta$ are the radiobiological parameters without HT. To assess the behaviour of the temperature-dependent rate $k(T)$, we calculated $TER_{\alpha (\text{ or }\beta)}-1$ for every data point to compare with the result of eq. \ref{TF2}, which was rearranged for this purpose as follows:

\begin{equation}\label{TF3}
\text{TER}-1=atk(T)=a't\:e^{b(T-T_g)}.
\end{equation}

\begin{figure*}
	\centering
	\includegraphics[scale=0.29]{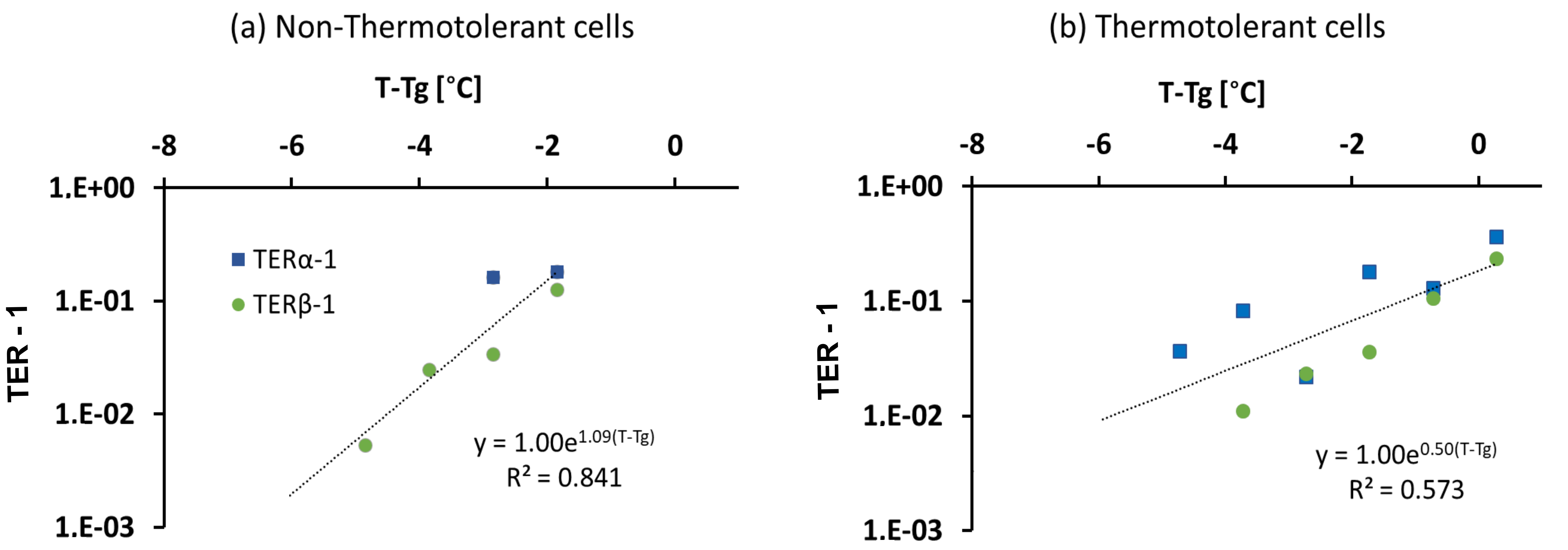}
	\caption{Thermal enhancement $(TER-1)$ as function of the relative temperature $(T-T_g)$ for M8013 mouse mammary carcinoma cells \textit{in vitro} \cite{Havemann}. (a) Thermotolerant modification of the cell line and (b) Non-thermotolerant cells. Vertical axes displayed in logarithmic scale. The lines are exponential fittings of the $TER_{\alpha}-1$ and $TER_{\beta}-1$ points together.}
	\label{fig3}
\end{figure*}

The results shown in Fig.\ref{fig3} display an exponential dependency of $k(T)$ with temperature in both cases - as predicted by our model. The model parameters ($a'$, $b$, $T_g$) and the coefficients of determination are presented in Table \ref{tab2}. Notably, the melting temperatures are quite similar for the two sublines, but the main difference comes from the slope of the calorimetry function $b=B/2k_B$, reflecting a possible slower denaturation of cellular proteins in the thermotolerant subline in response to heat. The parameters were adjusted for all TER values obtained from Eqs. \ref{terab}. However, it must be noted that the authors of this study reported problematic deviations in the measurements of $\alpha$ \cite{Havemann}, which may explain the low coefficients of determination shown in Table \ref{tab2}. When the adjustment is made using only the $TER_{\beta}$ experimental points, it improves to $R^2$ = 0.986 and 0.951 for thermotolerant and non-thermotolerant M8013-cells, respectively.

%\twocolumngrid
\begin{table}[h]
\centering
%\begin{longtable}[H] {|p{2cm}|p{2.8cm}|p{2.8cm}|p{2.8cm}|}
\caption{Parameters of the TER model Eq.\ref{TF3}, obtained from M8013 mouse mammary carcinoma \textit{in vitro} \cite{Havemann}}
\begin{tabular}{| c | c | c | c | c | c |}
\hline
\textbf{Cell model}     &   \textbf{$a'$}   &\textbf{$b$}   &   \textbf{$T_g$[$^\circ$C]}   &   \textbf{$R^2$} \\ \hline
M8013 Thermotolerant & 1.00 & 0.50 & 46.34 & 0.573 \\
M8013 Non-thermotolerant  & 1.00 & 1.09 & 46.47 & 0.841 \\ \hline
\end{tabular}
\label{tab2}
\end{table}

%All model parameters were obtained from fitting of the radiation response curves, given the existence of complete data sets where RT is combined with HT at different temperatures and treatment times. 
We must stress that this linear model is valid in the regime of non-ablative HT (40-46 $^{\circ}$C), %but is likely that for higher temperatures the linear effect is lost, adopting a different functional form. Notwithstanding, mild temperatures are 
which is used for radiosensitisation purposes at which heat-induced damage is primarily sublethal \cite{Lepock2,Jung}.\\ 

As can be seen in Tables \ref{tab1} and \ref{tab2}, the parameter related to the number of cells ($a'= ac $) is set to one for all the tumour cell models. Doing so, one could speculate that the slope in equation \ref{TF2} is completely modelled by the exponential factor which is solely a function of the thermodynamic quantities describing the heat capacity, namely the melting point $T_g$ and the slope of the calorimetry peak. Remarkably, this calorimetry peak is also very similar for the three non-thermotolerant tumour cell models, but lower for the thermotolerant one. This result may indicate the calorimetry peak as a possible marker for cellular thermotolerance. Moreover, we obtained different melting points  for CHO (\textit{in-vitro}), M8013 (\textit{in-vitro}), and C3H (\textit{in-vivo}), which are in the range of 46-49 $^{\circ}$C. %the melting temperatures are different for CHO (\textit{in-vitro}), M8013 (\textit{in-vitro}), and C3H (\textit{in-vivo}). 
This result is consistent with the findings from Lepock and collaborators that show different melting points for distinct cell types \cite{Lepock2} in that temperature range. %However, the other parameters of the model turn out to be similar for the three data sets. %The parameter related to the number of cells ($a'= ac $) is equal to one for these very different tumour cell models. Despite the current lack of experimental data, i.e. for other tumour models, one could speculate that the slope in equation \ref{TF2} is completely modelled by the exponential factor. This factor is solely a function of the thermodynamic quantities describing the heat capacity, namely the melting point $T_g$ and the slope of the calorimetry peak. Remarkably, this calorimetry peak is also very similar for the three non-thermotolerant tumour cell models, but lower for the thermotolerant one. This result may indicate the calorimetry peak as a possible marker for cellular thermotolerance. 
This should be verified by meticulous future experimental work. Calorimetry assays together with systematic TER measurements in various tumour cell models will be particularly relevant in this context, and can lead to a considerable reduction in the number of adjustable parameters. Nonetheless, our model already quite well predicts and explains from thermodynamic principles the modulation of radioresponse caused by HT treatment with at most three adjustable parameters. %Two of these three parameters ($B$ and $T_g$)come from the thermodynamic properties of the cellular proteins, which can be assessed through calorimetry measurements.}

\section{Conclusions and perspectives} \label{Con}

In this work, we have proposed the enhancement of radiotherapy by hyperthermia as the result of the increased vulnerability of a cell. It is achieved by the accumulation of sublethal damage either repairable or not. In our model, the radiosensitisation, quantified by TER, is proportional to the energy invested to induce this damage. We propose that in the range of mild HT, this energy and therefore the synergistic effect measured by TER, is a rate-dependent process that increases linearly with the time of heat application. We found the rate of the process to increase exponentially with temperature as a result of the chemical reactions involved in the protein denaturation process, which is induced by HT. This model offers a thermodynamics based approach to explain experimental results previously obtained in different studies. \\

Simultaneous thermoradiotherapy supposedly leads to higher TERs than sequential HT + RT modalities \cite{van_Leeuwen20172,van_Leeuwen2017,Overgaard2013,Overgaard}. This is true for both tumour and normal tissue. Consequently, normal tissue has to be spared to achieve therapeutic gain, which requires precise application of heat to the tumour. This has remained challenging, due to blood flow, re-oxygenation and heat dispersion, despite the fact that precise real-time temperature control and monitoring techniques are already in place \cite{Crezee}.\\ 

Today, various technologies, which are based on ultrasound, micro or electromagnetic waves, are available for simultaneous HT+RT treatment in the clinical setting and have for example been applied to treat breast cancer (reviewed in \cite{Moros}) and different types of superficial malignancies \cite{Myerson1999}. Others are under further development and/or envisioned for the treatment of different types of surface and deep-tissue tumours (reviewed in \cite{Crezee}). In this vein, new methodologies such as nanoparticles-HT and the design of precise simultaneous applicators \cite{Moros,Oberacker,Kosterev} have opened attractive prospects for implementing simultaneous thermoradiotherapy in standard clinical practice. Comprehensive mathematical modelling to better predict treatment outcome - as presented in the present article - will critically contribute to this process towards clinical routine.\\

Our model for simultaneous treatment is based on the modulation of the radiobiological parameters of the LQ-model. It is suitable to reproduce clonogenic survival curves, and very well reproduces TERs of \textit{in vitro} and \textit{in vivo} experiments. However, translation into more relevant clinical outcomes is still required, i.e. tumour control probabilities or control doses, where \comillas{disease control} means the long-term extinction of replication-competent tumour cells \textit{in vivo} after completion of the treatment \cite{Zaider}. This translation is usually done by means of simple logistic functions. Since the extraction of systematic radiobiological parameters from humans is difficult, even unfeasible for different treatment times and temperatures, the underlying parameters often come from classical 2D cell cultures. However, these have been found to insufficiently estimate radiation response \cite{Hillen,Naqa}. Indeed, more accurate translations require more elaborated approaches to reflect the treatment response in a more realistic and complex \textit{in vivo}-like environment. Examples of factors that need to be considered include cell-cell and cell-matrix interactions, oxygen distributions, proliferative activity and cell cycle progression in a 3-D cellular context. We therefore intend to combine the present theoretical findings with cellular automaton simulations in a next step, to model the treatment outcome in 3D tumour cell models such as \textit{in-silico} multicellular tumour spheroids.\\ 

The present study constitutes a crucial initial step for implementing more complex scenarios with respect to I. heterogeneous tumour cellular environment and micromileau in a 3D geometry and II. treatment modality, such as sequential application of HT and RT treatment, in which the thermodynamics of reversible/repairable effects needs to be included. Notably, sequential treatments are simpler in clinical handling and thus of high practical relevance. Our work paves the ground for a more elaborate unified model, which is in the focus of our ongoing work, with the aim to describe the individual treatments and their sequential application from common general principles.

\section{Methods} \label{Met}

\subsection{Thermal enhancement of radiotherapy}

When a certain radiation dose $D_R$ is applied to a set of living cells, the reduction rate is proportional to the number of cells at the time of the treatment

\begin{equation}
\frac{dN}{dD_R}=-\alpha N.
\end{equation}

Therefore, the direct transition from the \textit{alive} state of the cell to the \textit{dead} state obeys an exponential behavior $S=e^{-\alpha D_R}$, where $S=N/N_0$ is the survival fraction, and $\alpha$ defines the transition rate per dose, as depicted in Fig.\ref{fig1}(a) \cite{RBbook}. If the killing effect is composed of a direct killing path $\alpha$, and a secondary path composed of two or more stages of sublethal damage (SLD) accumulation, the logarithmic survival curve acquires a shoulder, as depicted in Fig.\ref{fig1}(b). In the particular case of the LQ-model, the exponent has a linear and a quadratic contribution, corresponding to direct killing and SLD accumulation respectively

\begin{equation}\label{LQ}
-\ln(S)=\alpha D_R +\beta D_R^2 \text{ .} \\
\end{equation}

\noindent The LQ-model was originally employed as an empirical approach \cite{Lea}; later Chadwick and Leenhouts \cite{Chadwick} proposed a molecular interpretation based on a statistical approach. In their interpretation cell death occurs due to double-strand breaks (DSB) of DNA, such that $\alpha$ and $\beta$ account for the probability of producing irreparable DSB as a consequence of one or two photon/particle hits, respectively. 
As a consequence of the sensitisation effect of HT, the radiation dose $D_{R+H}$ required to produce the same surviving fraction is reduced. This reduction implies in Eq.\ref{LQ}, that $\alpha$ and $\beta$ are increased to $\alpha^*$ and $\beta^*$ (in order to obtain the same therapeutic outcome), assessing the increased sensitivity of the cells as a consequence of heat 

\begin{equation}\label{LQ1}
-\ln(S)=\alpha^* D_{R+H} +\beta^* D_{R+H}^2 \text{ .} \\
\end{equation}

This radiosensitising effect of hyperthermia is quantified by the \textit{thermal enhancement ratio} TER. It is defined as the ratio between the radiation dose required to achieve a specific endpoint with ionizing radiation alone ($D_R$), and the radiation dose resulting in the same endpoint value when combined with hyperthermia ($D_{R+H}$):

\begin{equation}
\text{TER}=\frac{D_R}{D_{R+H}},
\end{equation}

\noindent with $D_R>0$ and $D_{R+H}>0$. The new linear and quadratic coefficients of the LQ-model are obtained by replacing $D_R$ with $D_{R+H}\text{TER}$ in Eq.\ref{LQ}:

\begin{equation}\label{BE}
-\ln(S)=\alpha \text{TER} D_{R+H} +\beta (\text{TER})^2 D_{R+H}^2 \text{ .} \\
\end{equation}

\noindent Comparing equations \ref{LQ1} and \ref{BE} shows how the radiobiological parameters are effectively rescaled by hyperthermia to $\alpha^*= \alpha\text{TER}$ and $\beta^*=\beta\text{TER}^2$. Notably TER has a stronger effect on $\beta^*$, bending the survival curves to lower survival values, in accordance with previous empirical data from experimental and clinical values studies \cite{Franken,Datta}, bending the survival curves to lower survival values. %The linear component a determines the initial slope of the cell survival curve and the effect of single ionization events at low RT doses whilst b, the quadratic component, represents the cell survival after two or more chromosomal lesions
We propose a model for TER as a function of HT parameters, namely temperature and time, which is incorporated into the LQ-model to predict the survival probability of RT combined with mild HT. As detailed in the results section, TER is assumed to be proportional to the energy absorbed in the transition from the live state (A) to the more vulnerable state (A') $E_{A \rightarrow A'}$, which in turn is defined as a rate-limited process

\begin{equation}\label{TER}
TER \propto E_{A \rightarrow A'} = c_1 + c_2 k(T)t \text{ ,}
\end{equation}

\noindent where $c_1$ is the baseline of TER, and $c_2$ accounts for the cell-line specific radio- and thermal sensitivity. In the absence of hyperthermia $TER=1$, resulting in $c_1=1$. The transition rate from (A) to (A') $k(T)$ is modelled assuming protein denaturation as the mechanism responsible for heat-induced cell damage, as described in the next section.

%\subsection{The oxygen effect}

\subsection{Temperature dependence of the transition rate}

The temperature dependency of the transition rate $k(T)$ is modelled by means of the Eyring's \textit{transition state theory} \cite{Eyring}:

\begin{equation}\label{rate}
k(T)=\left( \frac{K_B}{h_p}\right) T e^{-\frac{\Delta G(T)}{k_B T}},
\end{equation}

\noindent where $k_B$ and $h_p$ are the Boltzmann and Planck constants, respectively, and $T$ is the temperature in Kelvin. We next introduce a suitable model for the change in Gibbs energy $\Delta G(T)$ consistent with protein denaturation.\\

All conformation changes during protein denaturation arise from the competition between formation and breakage of chemical bonds. Protein denaturation becomes thermodynamically more favourable with increasing temperature. The dynamics of protein bonds is quantified by the \textit{standard heat of reaction} $\Delta H_0$ and the \textit{thermal work function} $\Delta W(T)$ respectively. We model the mixture of proteins sensitive to hyperthermia, as an average equivalent protein. Its overall heat capacity changes as a result of the state changes of individual proteins within the mixture. All the $\Delta$ symbols refer to changes in the thermodynamic properties of this \comillas{equivalent} protein, before and after the transformation.\\

The energy source for bonds to break is the \textit{thermal content} $\int_{T_0}^T C_p(T')dT'$, which refers to the heat absorbed during the process of protein unfolding while temperature increases. Only a part of the absorbed heat can be converted into bond-breaking work, as restricted by the second law of thermodynamics. The unused proportion of thermal content goes into entropy --the thermal work-- and is proportional to the absorbed heat and the relative temperature increment. The expressions for enthalpy and work content read as \cite{Benzinger}:

\begin{align}
\nonumber \Delta H(T)&= \Delta H_0+\int_{T_0}^T C_p(T')dT' \text{,}\\
 \Delta W(T)&= \int_{T_0}^T C_p\frac{(T-T')}{T'} dT',
\end{align}

\noindent where $\Delta H$ is the \textit{enthalpy} of the reaction, containing the bond forming energy $\Delta H_0$ and the sum of isothermal transfers of heat $\int_{T_0}^T C_p(T')dT'$. Here $C_p(T)$ is the \textit{heat capacity}, which might vary with temperature, according to the third law of thermodynamics. The net driving energy is then given by the Gibbs free energy 

\begin{equation}
 \Delta G(T)= \Delta H_0-\Delta W(T)=\Delta H-T \Delta S(T) \text{,}
\end{equation}

\vspace{0.3cm}
\noindent where $\Delta S(T)=\int_0^T \frac{C_p(T')}{T'}dT'=\Delta S_0+\int_{T_0}^T \frac{C_p(T')}{T'}dT'$ is the entropy change, with $\Delta S_0$ as reference value. Accordingly, the Gibbs energy is expressed as

\begin{equation}\label{GE} 
\Delta G(T)= \Delta G_0 + \int^T_{T_0} dT' C_p(T')\left [1-\frac{T}{T'}\right] \text{ ,}
\end{equation}

\noindent where $\Delta G_0= \Delta H_0-T \Delta S_0$. The reference temperature can be chosen so that $\Delta H(T_0=T_h)=0$, $\Delta S(T_0=T_S)=0$, or $\Delta G(T_0=T_g)=0$. $T_g$ and $T_s$ are of particular interest since they define the melting and maximal stability temperatures of the protein, respectively. When bond formation and breakage reach a balanced state ($\Delta G(T_g)=0$) the reaction does not progress anymore. The \textit{melting temperature} is defined as the temperature at which the half of the proteins are denatured \cite{Bischof}. Due to the importance of protein denaturation, the melting point is used as the reference temperature from now on.\\

The next challenge is to model the heat capacity in aqueous solutions above physiological temperatures. The heat capacity is expected to increase with temperature before approaching the vicinity of the melting point, as a result of ongoing protein reconfigurations. Beyond the transition, exothermic co-aggregations of proteins occur and $C_p$ is expected to decrease due to the reduced degrees of freedom of more rigid proteins. With these arguments, we propose to consider the next order by introducing the heat capacity change as a linear function of $(T-T_g)$ \cite{Prabhu}, $C_p(T)=A -B\vert T-T_g \vert$, which is the same as $C_p(T)=A +B(T-T_g)$ for $T \leq T_g$, leading to

\begin{equation}
\Delta G(T)= \Delta G_c -\frac{B}{2}(T^2-T_g^2)+BT T_g\ln \left(\frac{T}{Tg}\right).
\label{GE2} 
\end{equation}

\noindent Here $\Delta G_c$, is the usual Gibbs energy resulting from the assumption of constant heat capacity change. By introducing Eq.\ref{GE2} in Eq.\ref{rate}, the transition rate for denaturation becomes

\begin{equation}\label{rate2}
k(T)=\left( \frac{K_BT}{h_p}\right) e^{-\frac{\Delta G_c}{K_B T}}
e^{\frac{B}{2K_B}(T-T_g) \left( 1+\frac{Tg}{T} \right)}.
\end{equation}

\noindent where the last term in Eq.\ref{GE2} should vanish, because $T_g/T$ is about one in the Kelvin scale for the hyperthermia temperature range (40-50$^{\circ}$C). The first two factors of Eq.\ref{rate2} slightly change ($\sim \pm 2.5\%$) in these regimes, and then the transition rate is dominated by the exponential behavior. Based on these considerations, $k(T)$ can be described as

\begin{equation}\label{rate3}
k(T) \approx c\: e^{b(T-T_g)},
\end{equation}

\noindent with $c=\left( \frac{K_BT}{h_p}\right)  e^{-\frac{\Delta G_c}{K_B T}}$ and $b=\frac{B}{2K_B}$ as --cell dependent-- adjustable parameters of the model.\\

\textbf{Acknowledgements}
This work was supported by the German Federal Ministry of Education and Research (BMBF; 03Z1N512). The authors would like to thank Dario Egloff for very interesting discussions and comments.\\

\textbf{Authors’ contribution statement}
A.D.M. conceived the presented idea and developed the theory. A.D.M., S.M., J.B., and J.K. performed the computations and analyzed the results. D.M. and L.K. supervised the findings of this work. All authors contributed to the interpretation of the results and designed, wrote and discussed the manuscript.\\



\begin{thebibliography}{60}%
\makeatletter
\providecommand \@ifxundefined [1]{%
 \@ifx{#1\undefined}
}%
\providecommand \@ifnum [1]{%
 \ifnum #1\expandafter \@firstoftwo
 \else \expandafter \@secondoftwo
 \fi
}%
\providecommand \@ifx [1]{%
 \ifx #1\expandafter \@firstoftwo
 \else \expandafter \@secondoftwo
 \fi
}%
\providecommand \natexlab [1]{#1}%
\providecommand \enquote  [1]{``#1''}%
\providecommand \bibnamefont  [1]{#1}%
\providecommand \bibfnamefont [1]{#1}%
\providecommand \citenamefont [1]{#1}%
\providecommand \href@noop [0]{\@secondoftwo}%
\providecommand \href [0]{\begingroup \@sanitize@url \@href}%
\providecommand \@href[1]{\@@startlink{#1}\@@href}%
\providecommand \@@href[1]{\endgroup#1\@@endlink}%
\providecommand \@sanitize@url [0]{\catcode `\\12\catcode `\$12\catcode
  `\&12\catcode `\#12\catcode `\^12\catcode `\_12\catcode `\%12\relax}%
\providecommand \@@startlink[1]{}%
\providecommand \@@endlink[0]{}%
\providecommand \url  [0]{\begingroup\@sanitize@url \@url }%
\providecommand \@url [1]{\endgroup\@href {#1}{\urlprefix }}%
\providecommand \urlprefix  [0]{URL }%
\providecommand \Eprint [0]{\href }%
\providecommand \doibase [0]{http://dx.doi.org/}%
\providecommand \selectlanguage [0]{\@gobble}%
\providecommand \bibinfo  [0]{\@secondoftwo}%
\providecommand \bibfield  [0]{\@secondoftwo}%
\providecommand \translation [1]{[#1]}%
\providecommand \BibitemOpen [0]{}%
\providecommand \bibitemStop [0]{}%
\providecommand \bibitemNoStop [0]{.\EOS\space}%
\providecommand \EOS [0]{\spacefactor3000\relax}%
\providecommand \BibitemShut  [1]{\csname bibitem#1\endcsname}%
\let\auto@bib@innerbib\@empty
%</preamble>
\bibitem [{\citenamefont {Bray}\ \emph {et~al.}(2018)\citenamefont {Bray},
  \citenamefont {Ferlay}, \citenamefont {Soerjomataram}, \citenamefont
  {Siegel}, \citenamefont {Torre},\ and\ \citenamefont {Jemal}}]{stats}%
  \BibitemOpen
  \bibfield  {author} {\bibinfo {author} {\bibfnamefont {F.}~\bibnamefont
  {Bray}}, \bibinfo {author} {\bibfnamefont {J.}~\bibnamefont {Ferlay}},
  \bibinfo {author} {\bibfnamefont {I.}~\bibnamefont {Soerjomataram}}, \bibinfo
  {author} {\bibfnamefont {R.~L.}\ \bibnamefont {Siegel}}, \bibinfo {author}
  {\bibfnamefont {L.~A.}\ \bibnamefont {Torre}}, \ and\ \bibinfo {author}
  {\bibfnamefont {A.}~\bibnamefont {Jemal}},\ }\href {\doibase
  10.3322/caac.21492} {\bibfield  {journal} {\bibinfo  {journal} {CA: A Cancer
  Journal for Clinicians}\ }\textbf {\bibinfo {volume} {68}},\ \bibinfo {pages}
  {394} (\bibinfo {year} {2018})}\BibitemShut {NoStop}%
\bibitem [{\citenamefont {Yagawa}\ \emph {et~al.}(2017)\citenamefont {Yagawa},
  \citenamefont {Tanigawa}, \citenamefont {Kobayashi},\ and\ \citenamefont
  {Yamamoto}}]{Yagawa}%
  \BibitemOpen
  \bibfield  {author} {\bibinfo {author} {\bibfnamefont {Y.}~\bibnamefont
  {Yagawa}}, \bibinfo {author} {\bibfnamefont {K.}~\bibnamefont {Tanigawa}},
  \bibinfo {author} {\bibfnamefont {Y.}~\bibnamefont {Kobayashi}}, \ and\
  \bibinfo {author} {\bibfnamefont {M.}~\bibnamefont {Yamamoto}},\ }\href
  {\doibase 10.20517/2394-4722.2017.35} {\bibfield  {journal} {\bibinfo
  {journal} {Journal of Cancer Metastasis and Treatment}\ }\textbf {\bibinfo
  {volume} {13}},\ \bibinfo {pages} {218} (\bibinfo {year} {2017})}\BibitemShut
  {NoStop}%
\bibitem [{\citenamefont {Cheng}\ \emph {et~al.}(2019)\citenamefont {Cheng},
  \citenamefont {Weng}, \citenamefont {Yu}, \citenamefont {Zhu}, \citenamefont
  {Yang},\ and\ \citenamefont {Yuan}}]{Cheng}%
  \BibitemOpen
  \bibfield  {author} {\bibinfo {author} {\bibfnamefont {Y.}~\bibnamefont
  {Cheng}}, \bibinfo {author} {\bibfnamefont {S.}~\bibnamefont {Weng}},
  \bibinfo {author} {\bibfnamefont {L.}~\bibnamefont {Yu}}, \bibinfo {author}
  {\bibfnamefont {N.}~\bibnamefont {Zhu}}, \bibinfo {author} {\bibfnamefont
  {M.}~\bibnamefont {Yang}}, \ and\ \bibinfo {author} {\bibfnamefont
  {Y.}~\bibnamefont {Yuan}},\ }\href {\doibase 10.1177/1534735419876345}
  {\bibfield  {journal} {\bibinfo  {journal} {Integrative cancer therapies}\
  }\textbf {\bibinfo {volume} {18}},\ \bibinfo {pages} {1} (\bibinfo {year}
  {2019})}\BibitemShut {NoStop}%
\bibitem [{\citenamefont {Paulides}\ \emph {et~al.}(2016)\citenamefont
  {Paulides}, \citenamefont {Verduijn},\ and\ \citenamefont
  {Van~Holthe}}]{Paulides}%
  \BibitemOpen
  \bibfield  {author} {\bibinfo {author} {\bibfnamefont {M.~M.}\ \bibnamefont
  {Paulides}}, \bibinfo {author} {\bibfnamefont {G.~M.}\ \bibnamefont
  {Verduijn}}, \ and\ \bibinfo {author} {\bibfnamefont {N.}~\bibnamefont
  {Van~Holthe}},\ }\href {\doibase 10.1186/s13014-016-0588-8} {\bibfield
  {journal} {\bibinfo  {journal} {Radiation Oncology}\ }\textbf {\bibinfo
  {volume} {11}},\ \bibinfo {pages} {1} (\bibinfo {year} {2016})}\BibitemShut
  {NoStop}%
\bibitem [{\citenamefont {Kang}\ \emph {et~al.}(2020)\citenamefont {Kang},
  \citenamefont {Kim}, \citenamefont {Shin}, \citenamefont {Han}, \citenamefont
  {Won}, \citenamefont {Her}, \citenamefont {Park},\ and\ \citenamefont
  {Oh}}]{Kang}%
  \BibitemOpen
  \bibfield  {author} {\bibinfo {author} {\bibfnamefont {J.~K.}\ \bibnamefont
  {Kang}}, \bibinfo {author} {\bibfnamefont {J.~C.}\ \bibnamefont {Kim}},
  \bibinfo {author} {\bibfnamefont {Y.}~\bibnamefont {Shin}}, \bibinfo {author}
  {\bibfnamefont {S.~M.}\ \bibnamefont {Han}}, \bibinfo {author} {\bibfnamefont
  {W.~R.}\ \bibnamefont {Won}}, \bibinfo {author} {\bibfnamefont
  {J.}~\bibnamefont {Her}}, \bibinfo {author} {\bibfnamefont {J.~Y.}\
  \bibnamefont {Park}}, \ and\ \bibinfo {author} {\bibfnamefont {K.~T.}\
  \bibnamefont {Oh}},\ }\href {\doibase 10.1007/s12272-020-01206-5} {\bibfield
  {journal} {\bibinfo  {journal} {Archives of Pharmacal Research}\ }\textbf
  {\bibinfo {volume} {43}},\ \bibinfo {pages} {46} (\bibinfo {year}
  {2020})}\BibitemShut {NoStop}%
\bibitem [{\citenamefont {Soares}\ \emph {et~al.}(2011)\citenamefont {Soares},
  \citenamefont {Ferreira}, \citenamefont {Igreja}, \citenamefont {Novo},\ and\
  \citenamefont {Borges}}]{Soares}%
  \BibitemOpen
  \bibfield  {author} {\bibinfo {author} {\bibfnamefont {P.~I.~P.}\
  \bibnamefont {Soares}}, \bibinfo {author} {\bibfnamefont {I.~M.~M.}\
  \bibnamefont {Ferreira}}, \bibinfo {author} {\bibfnamefont {R.~A. G. B.~N.}\
  \bibnamefont {Igreja}}, \bibinfo {author} {\bibfnamefont {C.~M.~M.}\
  \bibnamefont {Novo}}, \ and\ \bibinfo {author} {\bibfnamefont {J.~P. M.~R.}\
  \bibnamefont {Borges}},\ }\href {\doibase 10.2174/157489212798358038}
  {\bibfield  {journal} {\bibinfo  {journal} {Recent patents on anti-cancer
  drug discovery}\ }\textbf {\bibinfo {volume} {7}},\ \bibinfo {pages} {64}
  (\bibinfo {year} {2011})}\BibitemShut {NoStop}%
\bibitem [{\citenamefont {Moros}\ \emph {et~al.}(2010)\citenamefont {Moros},
  \citenamefont {Peñagaricano}, \citenamefont {Novàk}, \citenamefont
  {Straube},\ and\ \citenamefont {Myerson}}]{Moros}%
  \BibitemOpen
  \bibfield  {author} {\bibinfo {author} {\bibfnamefont {E.~G.}\ \bibnamefont
  {Moros}}, \bibinfo {author} {\bibfnamefont {J.}~\bibnamefont
  {Peñagaricano}}, \bibinfo {author} {\bibfnamefont {P.}~\bibnamefont
  {Novàk}}, \bibinfo {author} {\bibfnamefont {W.~L.}\ \bibnamefont {Straube}},
  \ and\ \bibinfo {author} {\bibfnamefont {R.~J.}\ \bibnamefont {Myerson}},\
  }\href {\doibase 10.3109/02656736.2010.493915} {\bibfield  {journal}
  {\bibinfo  {journal} {International Journal of Hyperthermia}\ }\textbf
  {\bibinfo {volume} {26}},\ \bibinfo {pages} {699} (\bibinfo {year}
  {2010})}\BibitemShut {NoStop}%
\bibitem [{\citenamefont {Oberacker}\ \emph {et~al.}(2017)\citenamefont
  {Oberacker}, \citenamefont {Kuehne}, \citenamefont {Nadobny}, \citenamefont
  {Zschaeck}, \citenamefont {Weihrauch}, \citenamefont {Waiczies},
  \citenamefont {Ghadjar}, \citenamefont {Wust}, \citenamefont {Niendorf},\
  and\ \citenamefont {Winter}}]{Oberacker}%
  \BibitemOpen
  \bibfield  {author} {\bibinfo {author} {\bibfnamefont {E.}~\bibnamefont
  {Oberacker}}, \bibinfo {author} {\bibfnamefont {A.}~\bibnamefont {Kuehne}},
  \bibinfo {author} {\bibfnamefont {J.}~\bibnamefont {Nadobny}}, \bibinfo
  {author} {\bibfnamefont {S.}~\bibnamefont {Zschaeck}}, \bibinfo {author}
  {\bibfnamefont {M.}~\bibnamefont {Weihrauch}}, \bibinfo {author}
  {\bibfnamefont {H.}~\bibnamefont {Waiczies}}, \bibinfo {author}
  {\bibfnamefont {P.}~\bibnamefont {Ghadjar}}, \bibinfo {author} {\bibfnamefont
  {P.}~\bibnamefont {Wust}}, \bibinfo {author} {\bibfnamefont {T.}~\bibnamefont
  {Niendorf}}, \ and\ \bibinfo {author} {\bibfnamefont {L.}~\bibnamefont
  {Winter}},\ }\href {\doibase https://doi.org/10.1515/cdbme-2017-0100}
  {\bibfield  {journal} {\bibinfo  {journal} {Current Directions in Biomedical
  Engineering}\ }\textbf {\bibinfo {volume} {3}},\ \bibinfo {pages} {473 }
  (\bibinfo {year} {2017})}\BibitemShut {NoStop}%
\bibitem [{\citenamefont {Kosterev}\ \emph {et~al.}(2015)\citenamefont
  {Kosterev}, \citenamefont {Kramer-Ageev}, \citenamefont {Mazokhin},
  \citenamefont {van Rhoon},\ and\ \citenamefont {Crezee}}]{Kosterev}%
  \BibitemOpen
  \bibfield  {author} {\bibinfo {author} {\bibfnamefont {V.~V.}\ \bibnamefont
  {Kosterev}}, \bibinfo {author} {\bibfnamefont {E.~A.}\ \bibnamefont
  {Kramer-Ageev}}, \bibinfo {author} {\bibfnamefont {V.~N.}\ \bibnamefont
  {Mazokhin}}, \bibinfo {author} {\bibfnamefont {G.~C.}\ \bibnamefont {van
  Rhoon}}, \ and\ \bibinfo {author} {\bibfnamefont {J.}~\bibnamefont
  {Crezee}},\ }\href {\doibase 10.3109/02656736.2015.1026413} {\bibfield
  {journal} {\bibinfo  {journal} {International Journal of Hyperthermia}\
  }\textbf {\bibinfo {volume} {31}},\ \bibinfo {pages} {443} (\bibinfo {year}
  {2015})}\BibitemShut {NoStop}%
\bibitem [{CT()}]{CT}%
  \BibitemOpen
  \href@noop {} {\enquote {\bibinfo {title} {Clinicaltrials.gov},}\ }\bibinfo
  {howpublished} {\url{http://ClinicalTrials.gov}},\ \bibinfo {note} {accessed:
  2020-05-14}\BibitemShut {NoStop}%
\bibitem [{\citenamefont {Dracham}\ \emph {et~al.}(2018)\citenamefont
  {Dracham}, \citenamefont {Shankar},\ and\ \citenamefont {Madan}}]{Dracham}%
  \BibitemOpen
  \bibfield  {author} {\bibinfo {author} {\bibfnamefont {C.~B.}\ \bibnamefont
  {Dracham}}, \bibinfo {author} {\bibfnamefont {A.}~\bibnamefont {Shankar}}, \
  and\ \bibinfo {author} {\bibfnamefont {R.}~\bibnamefont {Madan}},\ }\href
  {\doibase 10.3857/roj.2018.00290} {\bibfield  {journal} {\bibinfo  {journal}
  {Radiat Oncol J}\ }\textbf {\bibinfo {volume} {36}},\ \bibinfo {pages} {85}
  (\bibinfo {year} {2018})}\BibitemShut {NoStop}%
\bibitem [{\citenamefont {Mei}\ \emph {et~al.}(2020)\citenamefont {Mei},
  \citenamefont {ten Cate}, \citenamefont {van Leeuwen}, \citenamefont
  {Rodermond}, \citenamefont {de~Leeuw}, \citenamefont {Dimitrakopoulou},
  \citenamefont {Stalpers}, \citenamefont {Crezee}, \citenamefont {Kok},
  \citenamefont {Franken},\ and\ \citenamefont {Oei}}]{Mei}%
  \BibitemOpen
  \bibfield  {author} {\bibinfo {author} {\bibfnamefont {X.}~\bibnamefont
  {Mei}}, \bibinfo {author} {\bibfnamefont {R.}~\bibnamefont {ten Cate}},
  \bibinfo {author} {\bibfnamefont {C.~M.}\ \bibnamefont {van Leeuwen}},
  \bibinfo {author} {\bibfnamefont {H.~M.}\ \bibnamefont {Rodermond}}, \bibinfo
  {author} {\bibfnamefont {L.}~\bibnamefont {de~Leeuw}}, \bibinfo {author}
  {\bibfnamefont {D.}~\bibnamefont {Dimitrakopoulou}}, \bibinfo {author}
  {\bibfnamefont {L.~J.~A.}\ \bibnamefont {Stalpers}}, \bibinfo {author}
  {\bibfnamefont {J.}~\bibnamefont {Crezee}}, \bibinfo {author} {\bibfnamefont
  {H.~P.}\ \bibnamefont {Kok}}, \bibinfo {author} {\bibfnamefont {N.~A.~P.}\
  \bibnamefont {Franken}}, \ and\ \bibinfo {author} {\bibfnamefont {A.~L.}\
  \bibnamefont {Oei}},\ }\href {\doibase 10.3390/cancers12030582} {\bibfield
  {journal} {\bibinfo  {journal} {Cancers}\ }\textbf {\bibinfo {volume} {12}}
  (\bibinfo {year} {2020}),\ 10.3390/cancers12030582}\BibitemShut {NoStop}%
\bibitem [{\citenamefont {Overgaard}(2013)}]{Overgaard2013}%
  \BibitemOpen
  \bibfield  {author} {\bibinfo {author} {\bibfnamefont {J.}~\bibnamefont
  {Overgaard}},\ }\href {\doibase 10.1016/j.radonc.2013.11.004} {\bibfield
  {journal} {\bibinfo  {journal} {Radiotherapy and Oncology}\ }\textbf
  {\bibinfo {volume} {109}},\ \bibinfo {pages} {185} (\bibinfo {year}
  {2013})}\BibitemShut {NoStop}%
\bibitem [{\citenamefont {Elming}\ \emph {et~al.}(2019)\citenamefont {Elming},
  \citenamefont {S{\o}rensen}, \citenamefont {Oei}, \citenamefont {Franken},
  \citenamefont {Crezee}, \citenamefont {Overgaard},\ and\ \citenamefont
  {Horsman}}]{Elming}%
  \BibitemOpen
  \bibfield  {author} {\bibinfo {author} {\bibfnamefont {P.~B.}\ \bibnamefont
  {Elming}}, \bibinfo {author} {\bibfnamefont {B.~S.}\ \bibnamefont
  {S{\o}rensen}}, \bibinfo {author} {\bibfnamefont {A.~L.}\ \bibnamefont
  {Oei}}, \bibinfo {author} {\bibfnamefont {N.~A.~P.}\ \bibnamefont {Franken}},
  \bibinfo {author} {\bibfnamefont {J.}~\bibnamefont {Crezee}}, \bibinfo
  {author} {\bibfnamefont {J.}~\bibnamefont {Overgaard}}, \ and\ \bibinfo
  {author} {\bibfnamefont {M.~R.}\ \bibnamefont {Horsman}},\ }\href {\doibase
  10.3390/cancers11010060} {\bibfield  {journal} {\bibinfo  {journal}
  {Cancers}\ }\textbf {\bibinfo {volume} {11}} (\bibinfo {year} {2019}),\
  10.3390/cancers11010060}\BibitemShut {NoStop}%
\bibitem [{\citenamefont {Datta}\ \emph {et~al.}(2014)\citenamefont {Datta},
  \citenamefont {Puric}, \citenamefont {Schneider}, \citenamefont {Weber},
  \citenamefont {Rogers},\ and\ \citenamefont {Bodis}}]{Carbon}%
  \BibitemOpen
  \bibfield  {author} {\bibinfo {author} {\bibfnamefont {N.~R.}\ \bibnamefont
  {Datta}}, \bibinfo {author} {\bibfnamefont {E.}~\bibnamefont {Puric}},
  \bibinfo {author} {\bibfnamefont {R.}~\bibnamefont {Schneider}}, \bibinfo
  {author} {\bibfnamefont {D.~C.}\ \bibnamefont {Weber}}, \bibinfo {author}
  {\bibfnamefont {S.}~\bibnamefont {Rogers}}, \ and\ \bibinfo {author}
  {\bibfnamefont {S.}~\bibnamefont {Bodis}},\ }\href {\doibase
  10.3109/02656736.2014.963703} {\bibfield  {journal} {\bibinfo  {journal}
  {International Journal of Hyperthermia}\ }\textbf {\bibinfo {volume} {30}},\
  \bibinfo {pages} {524} (\bibinfo {year} {2014})}\BibitemShut {NoStop}%
\bibitem [{\citenamefont {Datta}\ and\ \citenamefont
  {Bodis}(2019{\natexlab{a}})}]{Neutrons}%
  \BibitemOpen
  \bibfield  {author} {\bibinfo {author} {\bibfnamefont {N.~R.}\ \bibnamefont
  {Datta}}\ and\ \bibinfo {author} {\bibfnamefont {S.}~\bibnamefont {Bodis}},\
  }\href {\doibase 10.1080/02656736.2019.1679895} {\bibfield  {journal}
  {\bibinfo  {journal} {International Journal of Hyperthermia}\ }\textbf
  {\bibinfo {volume} {36}},\ \bibinfo {pages} {1072} (\bibinfo {year}
  {2019}{\natexlab{a}})}\BibitemShut {NoStop}%
\bibitem [{\citenamefont {van~der Zee}(2002)}]{vanderZee}%
  \BibitemOpen
  \bibfield  {author} {\bibinfo {author} {\bibfnamefont {J.}~\bibnamefont
  {van~der Zee}},\ }\href {\doibase 10.1093/annonc/mdf280} {\bibfield
  {journal} {\bibinfo  {journal} {Annals of Oncology}\ }\textbf {\bibinfo
  {volume} {13}},\ \bibinfo {pages} {1173} (\bibinfo {year}
  {2002})}\BibitemShut {NoStop}%
\bibitem [{\citenamefont {Datta}\ \emph
  {et~al.}(2016{\natexlab{a}})\citenamefont {Datta}, \citenamefont {Schneider},
  \citenamefont {Puric}, \citenamefont {Ahlhelm}, \citenamefont {Marder},
  \citenamefont {Bodis},\ and\ \citenamefont {Weber}}]{Datta2016}%
  \BibitemOpen
  \bibfield  {author} {\bibinfo {author} {\bibfnamefont {N.~R.}\ \bibnamefont
  {Datta}}, \bibinfo {author} {\bibfnamefont {R.}~\bibnamefont {Schneider}},
  \bibinfo {author} {\bibfnamefont {E.}~\bibnamefont {Puric}}, \bibinfo
  {author} {\bibfnamefont {F.~J.}\ \bibnamefont {Ahlhelm}}, \bibinfo {author}
  {\bibfnamefont {D.}~\bibnamefont {Marder}}, \bibinfo {author} {\bibfnamefont
  {S.}~\bibnamefont {Bodis}}, \ and\ \bibinfo {author} {\bibfnamefont {D.~C.}\
  \bibnamefont {Weber}},\ }\href {\doibase 10.14338/IJPT-16-00016.1} {\bibfield
   {journal} {\bibinfo  {journal} {International Journal of Particle Therapy}\
  }\textbf {\bibinfo {volume} {3}},\ \bibinfo {pages} {327} (\bibinfo {year}
  {2016}{\natexlab{a}})}\BibitemShut {NoStop}%
\bibitem [{\citenamefont {Hurwitz}\ \emph {et~al.}(2011)\citenamefont
  {Hurwitz}, \citenamefont {Hansen}, \citenamefont {Prokopios-Davos},
  \citenamefont {Manola}, \citenamefont {Wang}, \citenamefont {Bornstein},
  \citenamefont {Hynynen},\ and\ \citenamefont {Kaplan}}]{Hurwitz}%
  \BibitemOpen
  \bibfield  {author} {\bibinfo {author} {\bibfnamefont {M.~D.}\ \bibnamefont
  {Hurwitz}}, \bibinfo {author} {\bibfnamefont {J.~L.}\ \bibnamefont {Hansen}},
  \bibinfo {author} {\bibfnamefont {S.}~\bibnamefont {Prokopios-Davos}},
  \bibinfo {author} {\bibfnamefont {J.}~\bibnamefont {Manola}}, \bibinfo
  {author} {\bibfnamefont {Q.}~\bibnamefont {Wang}}, \bibinfo {author}
  {\bibfnamefont {B.~A.}\ \bibnamefont {Bornstein}}, \bibinfo {author}
  {\bibfnamefont {K.}~\bibnamefont {Hynynen}}, \ and\ \bibinfo {author}
  {\bibfnamefont {I.~D.}\ \bibnamefont {Kaplan}},\ }\href {\doibase
  10.1002/cncr.25619} {\bibfield  {journal} {\bibinfo  {journal} {Cancer}\
  }\textbf {\bibinfo {volume} {117}},\ \bibinfo {pages} {510} (\bibinfo {year}
  {2011})}\BibitemShut {NoStop}%
\bibitem [{\citenamefont {Datta}\ \emph
  {et~al.}(2016{\natexlab{b}})\citenamefont {Datta}, \citenamefont {Puric},
  \citenamefont {Klingbiel}, \citenamefont {Gomez},\ and\ \citenamefont
  {Bodis}}]{Datta20162}%
  \BibitemOpen
  \bibfield  {author} {\bibinfo {author} {\bibfnamefont {N.~R.}\ \bibnamefont
  {Datta}}, \bibinfo {author} {\bibfnamefont {E.}~\bibnamefont {Puric}},
  \bibinfo {author} {\bibfnamefont {D.}~\bibnamefont {Klingbiel}}, \bibinfo
  {author} {\bibfnamefont {S.}~\bibnamefont {Gomez}}, \ and\ \bibinfo {author}
  {\bibfnamefont {S.}~\bibnamefont {Bodis}},\ }\href {\doibase
  https://doi.org/10.1016/j.ijrobp.2015.12.361} {\bibfield  {journal} {\bibinfo
   {journal} {International Journal of Radiation Oncology Biology Physics}\
  }\textbf {\bibinfo {volume} {94}},\ \bibinfo {pages} {1073} (\bibinfo {year}
  {2016}{\natexlab{b}})}\BibitemShut {NoStop}%
\bibitem [{\citenamefont {Bakker}\ \emph {et~al.}(2019)\citenamefont {Bakker},
  \citenamefont {van~der Zee}, \citenamefont {van Tienhoven}, \citenamefont
  {Kok}, \citenamefont {Rasch},\ and\ \citenamefont {Crezee}}]{Bakker}%
  \BibitemOpen
  \bibfield  {author} {\bibinfo {author} {\bibfnamefont {A.}~\bibnamefont
  {Bakker}}, \bibinfo {author} {\bibfnamefont {J.}~\bibnamefont {van~der Zee}},
  \bibinfo {author} {\bibfnamefont {G.}~\bibnamefont {van Tienhoven}}, \bibinfo
  {author} {\bibfnamefont {H.~P.}\ \bibnamefont {Kok}}, \bibinfo {author}
  {\bibfnamefont {C.~R.~N.}\ \bibnamefont {Rasch}}, \ and\ \bibinfo {author}
  {\bibfnamefont {H.}~\bibnamefont {Crezee}},\ }\href {\doibase
  10.1080/02656736.2019.1665718} {\bibfield  {journal} {\bibinfo  {journal}
  {International Journal of Hyperthermia}\ }\textbf {\bibinfo {volume} {36}},\
  \bibinfo {pages} {1023} (\bibinfo {year} {2019})}\BibitemShut {NoStop}%
\bibitem [{\citenamefont {van Leeuwen}\ \emph {et~al.}(2017)\citenamefont {van
  Leeuwen}, \citenamefont {Oei}, \citenamefont {Chin}, \citenamefont {Crezee},
  \citenamefont {Bel}, \citenamefont {Westermann}, \citenamefont {Buist},
  \citenamefont {Franken}, \citenamefont {Stalpers},\ and\ \citenamefont
  {Kok}}]{van_Leeuwen20172}%
  \BibitemOpen
  \bibfield  {author} {\bibinfo {author} {\bibfnamefont {C.~M.}\ \bibnamefont
  {van Leeuwen}}, \bibinfo {author} {\bibfnamefont {A.~L.}\ \bibnamefont
  {Oei}}, \bibinfo {author} {\bibfnamefont {K.~W. T.~K.}\ \bibnamefont {Chin}},
  \bibinfo {author} {\bibfnamefont {J.}~\bibnamefont {Crezee}}, \bibinfo
  {author} {\bibfnamefont {A.}~\bibnamefont {Bel}}, \bibinfo {author}
  {\bibfnamefont {A.~M.}\ \bibnamefont {Westermann}}, \bibinfo {author}
  {\bibfnamefont {M.~R.}\ \bibnamefont {Buist}}, \bibinfo {author}
  {\bibfnamefont {N.~A.~P.}\ \bibnamefont {Franken}}, \bibinfo {author}
  {\bibfnamefont {L.~J.~A.}\ \bibnamefont {Stalpers}}, \ and\ \bibinfo {author}
  {\bibfnamefont {H.~P.}\ \bibnamefont {Kok}},\ }\href {\doibase
  10.1186/s13014-017-0813-0} {\bibfield  {journal} {\bibinfo  {journal}
  {Radiation Oncology}\ }\textbf {\bibinfo {volume} {12}},\ \bibinfo {pages}
  {1} (\bibinfo {year} {2017})}\BibitemShut {NoStop}%
\bibitem [{\citenamefont {van Leeuwen}\ \emph
  {et~al.}(2018{\natexlab{a}})\citenamefont {van Leeuwen}, \citenamefont
  {Crezee}, \citenamefont {Oei}, \citenamefont {Franken}, \citenamefont
  {Stalpers}, \citenamefont {Bel},\ and\ \citenamefont
  {Kok}}]{van_Leeuwen2017}%
  \BibitemOpen
  \bibfield  {author} {\bibinfo {author} {\bibfnamefont {C.~M.}\ \bibnamefont
  {van Leeuwen}}, \bibinfo {author} {\bibfnamefont {J.}~\bibnamefont {Crezee}},
  \bibinfo {author} {\bibfnamefont {A.~L.}\ \bibnamefont {Oei}}, \bibinfo
  {author} {\bibfnamefont {N.~A.~P.}\ \bibnamefont {Franken}}, \bibinfo
  {author} {\bibfnamefont {L.~J.~A.}\ \bibnamefont {Stalpers}}, \bibinfo
  {author} {\bibfnamefont {A.}~\bibnamefont {Bel}}, \ and\ \bibinfo {author}
  {\bibfnamefont {H.~P.}\ \bibnamefont {Kok}},\ }\href {\doibase
  10.1080/02656736.2018.1468930} {\bibfield  {journal} {\bibinfo  {journal}
  {International Journal of Hyperthermia}\ }\textbf {\bibinfo {volume} {34}},\
  \bibinfo {pages} {901} (\bibinfo {year} {2018}{\natexlab{a}})}\BibitemShut
  {NoStop}%
\bibitem [{\citenamefont {Overgaard}(1980)}]{Overgaard}%
  \BibitemOpen
  \bibfield  {author} {\bibinfo {author} {\bibfnamefont {J.}~\bibnamefont
  {Overgaard}},\ }\href {\doibase 10.1016/0360-3016(80)90008-5} {\bibfield
  {journal} {\bibinfo  {journal} {International Journal of Radiation
  Oncology*Biology*Physics}\ }\textbf {\bibinfo {volume} {6}},\ \bibinfo
  {pages} {1507} (\bibinfo {year} {1980})}\BibitemShut {NoStop}%
\bibitem [{\citenamefont {Spirou}\ \emph {et~al.}(2018)\citenamefont {Spirou},
  \citenamefont {Basini}, \citenamefont {Lascialfari}, \citenamefont
  {Sangregorio},\ and\ \citenamefont {Innocenti}}]{Spirou}%
  \BibitemOpen
  \bibfield  {author} {\bibinfo {author} {\bibfnamefont {S.~V.}\ \bibnamefont
  {Spirou}}, \bibinfo {author} {\bibfnamefont {M.}~\bibnamefont {Basini}},
  \bibinfo {author} {\bibfnamefont {A.}~\bibnamefont {Lascialfari}}, \bibinfo
  {author} {\bibfnamefont {C.}~\bibnamefont {Sangregorio}}, \ and\ \bibinfo
  {author} {\bibfnamefont {C.}~\bibnamefont {Innocenti}},\ }\href {\doibase
  doi: 10.3390/nano8060401} {\bibfield  {journal} {\bibinfo  {journal}
  {Nanomaterials}\ }\textbf {\bibinfo {volume} {8}},\ \bibinfo {pages} {1}
  (\bibinfo {year} {2018})}\BibitemShut {NoStop}%
\bibitem [{\citenamefont {Peeken}\ \emph {et~al.}(2017)\citenamefont {Peeken},
  \citenamefont {Vaupel},\ and\ \citenamefont {Combs}}]{Peeken}%
  \BibitemOpen
  \bibfield  {author} {\bibinfo {author} {\bibfnamefont {J.~C.}\ \bibnamefont
  {Peeken}}, \bibinfo {author} {\bibfnamefont {P.}~\bibnamefont {Vaupel}}, \
  and\ \bibinfo {author} {\bibfnamefont {S.~E.}\ \bibnamefont {Combs}},\ }\href
  {\doibase 10.3389/fonc.2017.00132} {\bibfield  {journal} {\bibinfo  {journal}
  {Frontiers in oncology}\ }\textbf {\bibinfo {volume} {7}},\ \bibinfo {pages}
  {1} (\bibinfo {year} {2017})}\BibitemShut {NoStop}%
\bibitem [{\citenamefont {Datta}\ \emph {et~al.}(2015)\citenamefont {Datta},
  \citenamefont {Gómez~Ordóñez}, \citenamefont {Gaipl}, \citenamefont
  {Paulides}, \citenamefont {Crezee}, \citenamefont {Gellermann}, \citenamefont
  {Marder}, \citenamefont {Puric},\ and\ \citenamefont {S.}}]{Datta2015}%
  \BibitemOpen
  \bibfield  {author} {\bibinfo {author} {\bibfnamefont {N.}~\bibnamefont
  {Datta}}, \bibinfo {author} {\bibfnamefont {S.}~\bibnamefont
  {Gómez~Ordóñez}}, \bibinfo {author} {\bibfnamefont {U.}~\bibnamefont
  {Gaipl}}, \bibinfo {author} {\bibfnamefont {M.}~\bibnamefont {Paulides}},
  \bibinfo {author} {\bibfnamefont {H.}~\bibnamefont {Crezee}}, \bibinfo
  {author} {\bibfnamefont {J.}~\bibnamefont {Gellermann}}, \bibinfo {author}
  {\bibfnamefont {D.}~\bibnamefont {Marder}}, \bibinfo {author} {\bibfnamefont
  {E.}~\bibnamefont {Puric}}, \ and\ \bibinfo {author} {\bibfnamefont
  {B.}~\bibnamefont {S.}},\ }\href {\doibase
  https://doi.org/10.1016/j.ctrv.2015.05.009} {\bibfield  {journal} {\bibinfo
  {journal} {Cancer treatments reviews}\ }\textbf {\bibinfo {volume} {41}},\
  \bibinfo {pages} {1} (\bibinfo {year} {2015})}\BibitemShut {NoStop}%
\bibitem [{\citenamefont {Varma}\ \emph {et~al.}(2012)\citenamefont {Varma},
  \citenamefont {Myerson}, \citenamefont {Moros}, \citenamefont {Taylor},
  \citenamefont {Straube},\ and\ \citenamefont {Zoberi}}]{Varma}%
  \BibitemOpen
  \bibfield  {author} {\bibinfo {author} {\bibfnamefont {S.}~\bibnamefont
  {Varma}}, \bibinfo {author} {\bibfnamefont {R.}~\bibnamefont {Myerson}},
  \bibinfo {author} {\bibfnamefont {E.}~\bibnamefont {Moros}}, \bibinfo
  {author} {\bibfnamefont {M.}~\bibnamefont {Taylor}}, \bibinfo {author}
  {\bibfnamefont {W.}~\bibnamefont {Straube}}, \ and\ \bibinfo {author}
  {\bibfnamefont {I.}~\bibnamefont {Zoberi}},\ }\href {\doibase
  10.3109/02656736.2012.705216} {\bibfield  {journal} {\bibinfo  {journal}
  {International Journal of Hyperthermia}\ }\textbf {\bibinfo {volume} {28}},\
  \bibinfo {pages} {583} (\bibinfo {year} {2012})}\BibitemShut {NoStop}%
\bibitem [{\citenamefont {Myerson}\ \emph {et~al.}(1999)\citenamefont
  {Myerson}, \citenamefont {Strause}, \citenamefont {Emami}, \citenamefont
  {Lee}, \citenamefont {Perez},\ and\ \citenamefont {E.}}]{Myerson1999}%
  \BibitemOpen
  \bibfield  {author} {\bibinfo {author} {\bibfnamefont {R.~J.}\ \bibnamefont
  {Myerson}}, \bibinfo {author} {\bibfnamefont {E.~G.}\ \bibnamefont {Strause},
  \bibfnamefont {W.~L.and~Moros}}, \bibinfo {author} {\bibfnamefont {B.~N.}\
  \bibnamefont {Emami}}, \bibinfo {author} {\bibfnamefont {H.~K.}\ \bibnamefont
  {Lee}}, \bibinfo {author} {\bibfnamefont {C.~A.}\ \bibnamefont {Perez}}, \
  and\ \bibinfo {author} {\bibfnamefont {T.~M.}\ \bibnamefont {E.}},\ }\href
  {\doibase 10.1080/026567399285639} {\bibfield  {journal} {\bibinfo  {journal}
  {International Journal of Hyperthermia}\ }\textbf {\bibinfo {volume} {15}},\
  \bibinfo {pages} {251} (\bibinfo {year} {1999})}\BibitemShut {NoStop}%
\bibitem [{\citenamefont {Chadwick}\ and\ \citenamefont
  {Leenhouts}(1973)}]{Chadwick}%
  \BibitemOpen
  \bibfield  {author} {\bibinfo {author} {\bibfnamefont {K.~H.}\ \bibnamefont
  {Chadwick}}\ and\ \bibinfo {author} {\bibfnamefont {H.~P.}\ \bibnamefont
  {Leenhouts}},\ }\href {\doibase 10.1088/0031-9155/18/1/007} {\bibfield
  {journal} {\bibinfo  {journal} {Physics in Medicine and Biology}\ }\textbf
  {\bibinfo {volume} {18}},\ \bibinfo {pages} {78} (\bibinfo {year}
  {1973})}\BibitemShut {NoStop}%
\bibitem [{\citenamefont {Datta}\ and\ \citenamefont
  {Bodis}(2019{\natexlab{b}})}]{Datta}%
  \BibitemOpen
  \bibfield  {author} {\bibinfo {author} {\bibfnamefont {N.~R.}\ \bibnamefont
  {Datta}}\ and\ \bibinfo {author} {\bibfnamefont {S.}~\bibnamefont {Bodis}},\
  }\href {\doibase 10.1016/j.radonc.2019.05.002} {\bibfield  {journal}
  {\bibinfo  {journal} {Radiotherapy and Oncology}\ }\textbf {\bibinfo {volume}
  {138}},\ \bibinfo {pages} {1} (\bibinfo {year}
  {2019}{\natexlab{b}})}\BibitemShut {NoStop}%
\bibitem [{\citenamefont {Crezee}\ \emph {et~al.}(2016)\citenamefont {Crezee},
  \citenamefont {van Leeuwen}, \citenamefont {Oei}, \citenamefont {Stalpers},
  \citenamefont {Bel}, \citenamefont {Franken},\ and\ \citenamefont
  {Kok}}]{Crezee}%
  \BibitemOpen
  \bibfield  {author} {\bibinfo {author} {\bibfnamefont {H.}~\bibnamefont
  {Crezee}}, \bibinfo {author} {\bibfnamefont {C.~M.}\ \bibnamefont {van
  Leeuwen}}, \bibinfo {author} {\bibfnamefont {A.~L.}\ \bibnamefont {Oei}},
  \bibinfo {author} {\bibfnamefont {L.~J.}\ \bibnamefont {Stalpers}}, \bibinfo
  {author} {\bibfnamefont {A.}~\bibnamefont {Bel}}, \bibinfo {author}
  {\bibfnamefont {N.~A.}\ \bibnamefont {Franken}}, \ and\ \bibinfo {author}
  {\bibfnamefont {H.~P.}\ \bibnamefont {Kok}},\ }\href {\doibase
  10.3109/02656736.2015.1110757} {\bibfield  {journal} {\bibinfo  {journal}
  {International Journal of Hyperthermia}\ }\textbf {\bibinfo {volume} {32}},\
  \bibinfo {pages} {41} (\bibinfo {year} {2016})}\BibitemShut {NoStop}%
\bibitem [{\citenamefont {Bodgi}\ \emph {et~al.}(2016)\citenamefont {Bodgi},
  \citenamefont {Canet}, \citenamefont {Pujo-Menjouet}, \citenamefont {Lesne},
  \citenamefont {Victor},\ and\ \citenamefont {Foray}}]{reviewRT}%
  \BibitemOpen
  \bibfield  {author} {\bibinfo {author} {\bibfnamefont {L.}~\bibnamefont
  {Bodgi}}, \bibinfo {author} {\bibfnamefont {A.}~\bibnamefont {Canet}},
  \bibinfo {author} {\bibfnamefont {L.}~\bibnamefont {Pujo-Menjouet}}, \bibinfo
  {author} {\bibfnamefont {A.}~\bibnamefont {Lesne}}, \bibinfo {author}
  {\bibfnamefont {J.-M.}\ \bibnamefont {Victor}}, \ and\ \bibinfo {author}
  {\bibfnamefont {N.}~\bibnamefont {Foray}},\ }\href {\doibase
  10.1016/j.jtbi.2016.01.018} {\bibfield  {journal} {\bibinfo  {journal}
  {Journal of Theoretical Biology}\ }\textbf {\bibinfo {volume} {394}},\
  \bibinfo {pages} {93} (\bibinfo {year} {2016})}\BibitemShut {NoStop}%
\bibitem [{\citenamefont {Joiner}\ and\ \citenamefont {van~der
  Kogel}(2009)}]{RBbook}%
  \BibitemOpen
  \bibfield  {author} {\bibinfo {author} {\bibfnamefont {C.}~\bibnamefont
  {Joiner}, \bibfnamefont {Michael}}\ and\ \bibinfo {author} {\bibfnamefont
  {A.~J.}\ \bibnamefont {van~der Kogel}},\ }\href@noop {} {\emph {\bibinfo
  {title} {Basic Clinical Radiobiology}}},\ \bibinfo {edition} {4th}\ ed.\
  (\bibinfo  {publisher} {CRC Press},\ \bibinfo {year} {2009})\BibitemShut
  {NoStop}%
\bibitem [{\citenamefont {Zaider}\ and\ \citenamefont {Hanin}(2011)}]{Zaider}%
  \BibitemOpen
  \bibfield  {author} {\bibinfo {author} {\bibfnamefont {M.}~\bibnamefont
  {Zaider}}\ and\ \bibinfo {author} {\bibfnamefont {L.}~\bibnamefont {Hanin}},\
  }\href {\doibase 10.1118/1.3521406} {\bibfield  {journal} {\bibinfo
  {journal} {Medical Physics}\ }\textbf {\bibinfo {volume} {38}},\ \bibinfo
  {pages} {574} (\bibinfo {year} {2011})}\BibitemShut {NoStop}%
\bibitem [{\citenamefont {Lepock}(2005)}]{Lepock1}%
  \BibitemOpen
  \bibfield  {author} {\bibinfo {author} {\bibfnamefont {J.~R.}\ \bibnamefont
  {Lepock}},\ }\href {\doibase 10.1080/02656730500307298} {\bibfield  {journal}
  {\bibinfo  {journal} {International Journal of Hyperthermia}\ }\textbf
  {\bibinfo {volume} {21}},\ \bibinfo {pages} {681} (\bibinfo {year}
  {2005})}\BibitemShut {NoStop}%
\bibitem [{\citenamefont {Lepock}\ \emph {et~al.}(1993)\citenamefont {Lepock},
  \citenamefont {Frey},\ and\ \citenamefont {Ritchie}}]{Lepock2}%
  \BibitemOpen
  \bibfield  {author} {\bibinfo {author} {\bibfnamefont {J.}~\bibnamefont
  {Lepock}}, \bibinfo {author} {\bibfnamefont {H.}~\bibnamefont {Frey}}, \ and\
  \bibinfo {author} {\bibfnamefont {K.}~\bibnamefont {Ritchie}},\ }\href
  {\doibase 10.1083/jcb.122.6.1267} {\bibfield  {journal} {\bibinfo  {journal}
  {Journal of Cell Biology}\ }\textbf {\bibinfo {volume} {122}},\ \bibinfo
  {pages} {1267} (\bibinfo {year} {1993})}\BibitemShut {NoStop}%
\bibitem [{\citenamefont {Grimm}\ \emph {et~al.}(2009)\citenamefont {Grimm},
  \citenamefont {Zynda},\ and\ \citenamefont {Repasky}}]{Grimm}%
  \BibitemOpen
  \bibfield  {author} {\bibinfo {author} {\bibfnamefont {M.}~\bibnamefont
  {Grimm}}, \bibinfo {author} {\bibfnamefont {E.}~\bibnamefont {Zynda}}, \ and\
  \bibinfo {author} {\bibfnamefont {E.}~\bibnamefont {Repasky}},\ }\enquote
  {\bibinfo {title} {Prokaryotic and eukaryotic heat shock proteins in
  infectious disease},}\ \ (\bibinfo  {publisher} {Springer,Dordrecht},\
  \bibinfo {year} {2009})\ Chap.\ \bibinfo {chapter} {Temperature Matters:
  Cellular Targets of Hyperthermia in Cancer Biology and
  Immunology}\BibitemShut {NoStop}%
\bibitem [{\citenamefont {Kampinga}\ and\ \citenamefont
  {Dikomey}(2001)}]{Kampinga3}%
  \BibitemOpen
  \bibfield  {author} {\bibinfo {author} {\bibfnamefont {H.~H.}\ \bibnamefont
  {Kampinga}}\ and\ \bibinfo {author} {\bibfnamefont {E.}~\bibnamefont
  {Dikomey}},\ }\href {\doibase 10.1080/09553000010024687} {\bibfield
  {journal} {\bibinfo  {journal} {International Journal of Radiation Biology}\
  }\textbf {\bibinfo {volume} {77}},\ \bibinfo {pages} {399} (\bibinfo {year}
  {2001})}\BibitemShut {NoStop}%
\bibitem [{\citenamefont {Pearce}(2013)}]{Pearce}%
  \BibitemOpen
  \bibfield  {author} {\bibinfo {author} {\bibfnamefont {J.~A.}\ \bibnamefont
  {Pearce}},\ }\href {\doibase 10.3109/02656736.2013.786140} {\bibfield
  {journal} {\bibinfo  {journal} {International Journal of Hyperthermia}\
  }\textbf {\bibinfo {volume} {29}},\ \bibinfo {pages} {262} (\bibinfo {year}
  {2013})}\BibitemShut {NoStop}%
\bibitem [{\citenamefont {Jung}(1991)}]{Jung}%
  \BibitemOpen
  \bibfield  {author} {\bibinfo {author} {\bibfnamefont {H.}~\bibnamefont
  {Jung}},\ }\href {\doibase 10.2307/3577936} {\bibfield  {journal} {\bibinfo
  {journal} {Radiation Research}\ }\textbf {\bibinfo {volume} {127}},\ \bibinfo
  {pages} {235} (\bibinfo {year} {1991})}\BibitemShut {NoStop}%
\bibitem [{\citenamefont {Oei}\ \emph {et~al.}(2015)\citenamefont {Oei},
  \citenamefont {Vriend}, \citenamefont {Crezee}, \citenamefont {Franken},\
  and\ \citenamefont {Krawczyk}}]{Oei}%
  \BibitemOpen
  \bibfield  {author} {\bibinfo {author} {\bibfnamefont {A.~L.}\ \bibnamefont
  {Oei}}, \bibinfo {author} {\bibfnamefont {L.~E.~M.}\ \bibnamefont {Vriend}},
  \bibinfo {author} {\bibfnamefont {J.}~\bibnamefont {Crezee}}, \bibinfo
  {author} {\bibfnamefont {N.~A.~P.}\ \bibnamefont {Franken}}, \ and\ \bibinfo
  {author} {\bibfnamefont {P.~M.}\ \bibnamefont {Krawczyk}},\ }\href {\doibase
  10.1186/s13014-015-0462-0} {\bibfield  {journal} {\bibinfo  {journal}
  {Radiation Oncology}\ }\textbf {\bibinfo {volume} {165}} (\bibinfo {year}
  {2015}),\ 10.1186/s13014-015-0462-0}\BibitemShut {NoStop}%
\bibitem [{\citenamefont {Br{\"{u}}ningk}\ \emph
  {et~al.}(2018{\natexlab{a}})\citenamefont {Br{\"{u}}ningk}, \citenamefont
  {Ijaz}, \citenamefont {Rivens}, \citenamefont {Nill}, \citenamefont {Haar},\
  and\ \citenamefont {Oelfke}}]{Bruningk1}%
  \BibitemOpen
  \bibfield  {author} {\bibinfo {author} {\bibfnamefont {S.~C.}\ \bibnamefont
  {Br{\"{u}}ningk}}, \bibinfo {author} {\bibfnamefont {J.}~\bibnamefont
  {Ijaz}}, \bibinfo {author} {\bibfnamefont {I.}~\bibnamefont {Rivens}},
  \bibinfo {author} {\bibfnamefont {S.}~\bibnamefont {Nill}}, \bibinfo {author}
  {\bibfnamefont {G.~t.}\ \bibnamefont {Haar}}, \ and\ \bibinfo {author}
  {\bibfnamefont {U.}~\bibnamefont {Oelfke}},\ }\href {\doibase
  10.1080/02656736.2017.1341059} {\bibfield  {journal} {\bibinfo  {journal}
  {International Journal of Hyperthermia}\ }\textbf {\bibinfo {volume} {34}},\
  \bibinfo {pages} {392} (\bibinfo {year} {2018}{\natexlab{a}})},\ \bibinfo
  {note} {pMID: 28641499}\BibitemShut {NoStop}%
\bibitem [{\citenamefont {Br{\"{u}}ningk}\ \emph
  {et~al.}(2018{\natexlab{b}})\citenamefont {Br{\"{u}}ningk}, \citenamefont
  {Powathil}, \citenamefont {Ziegenhein}, \citenamefont {Ijaz}, \citenamefont
  {Rivens}, \citenamefont {Nill}, \citenamefont {Chaplain}, \citenamefont
  {Oelfke},\ and\ \citenamefont {ter Haar}}]{Bruningk2}%
  \BibitemOpen
  \bibfield  {author} {\bibinfo {author} {\bibfnamefont {S.}~\bibnamefont
  {Br{\"{u}}ningk}}, \bibinfo {author} {\bibfnamefont {G.}~\bibnamefont
  {Powathil}}, \bibinfo {author} {\bibfnamefont {P.}~\bibnamefont
  {Ziegenhein}}, \bibinfo {author} {\bibfnamefont {J.}~\bibnamefont {Ijaz}},
  \bibinfo {author} {\bibfnamefont {I.}~\bibnamefont {Rivens}}, \bibinfo
  {author} {\bibfnamefont {S.}~\bibnamefont {Nill}}, \bibinfo {author}
  {\bibfnamefont {M.}~\bibnamefont {Chaplain}}, \bibinfo {author}
  {\bibfnamefont {U.}~\bibnamefont {Oelfke}}, \ and\ \bibinfo {author}
  {\bibfnamefont {G.}~\bibnamefont {ter Haar}},\ }\href {\doibase
  10.1098/rsif.2017.0681} {\bibfield  {journal} {\bibinfo  {journal} {Journal
  of The Royal Society Interface}\ }\textbf {\bibinfo {volume} {15}},\ \bibinfo
  {pages} {20170681} (\bibinfo {year} {2018}{\natexlab{b}})}\BibitemShut
  {NoStop}%
\bibitem [{\citenamefont {van Leeuwen}\ \emph
  {et~al.}(2018{\natexlab{b}})\citenamefont {van Leeuwen}, \citenamefont {Oei},
  \citenamefont {ten Cate}, \citenamefont {Franken}, \citenamefont {Bel},
  \citenamefont {Stalpers}, \citenamefont {Crezee},\ and\ \citenamefont
  {Kok}}]{vanLeeuwen1}%
  \BibitemOpen
  \bibfield  {author} {\bibinfo {author} {\bibfnamefont {C.~M.}\ \bibnamefont
  {van Leeuwen}}, \bibinfo {author} {\bibfnamefont {A.~L.}\ \bibnamefont
  {Oei}}, \bibinfo {author} {\bibfnamefont {R.}~\bibnamefont {ten Cate}},
  \bibinfo {author} {\bibfnamefont {N.~A.~P.}\ \bibnamefont {Franken}},
  \bibinfo {author} {\bibfnamefont {A.}~\bibnamefont {Bel}}, \bibinfo {author}
  {\bibfnamefont {L.~J.~A.}\ \bibnamefont {Stalpers}}, \bibinfo {author}
  {\bibfnamefont {J.}~\bibnamefont {Crezee}}, \ and\ \bibinfo {author}
  {\bibfnamefont {H.~P.}\ \bibnamefont {Kok}},\ }\href {\doibase
  10.1080/02656736.2017.1320812} {\bibfield  {journal} {\bibinfo  {journal}
  {International Journal of Hyperthermia}\ }\textbf {\bibinfo {volume} {34}},\
  \bibinfo {pages} {30} (\bibinfo {year} {2018}{\natexlab{b}})},\ \bibinfo
  {note} {pMID: 28540813}\BibitemShut {NoStop}%
\bibitem [{\citenamefont {Gillette}(1984)}]{Gillette}%
  \BibitemOpen
  \bibfield  {author} {\bibinfo {author} {\bibfnamefont {E.~L.}\ \bibnamefont
  {Gillette}},\ }\href
  {https://cancerres.aacrjournals.org/content/44/10_Supplement/4836s}
  {\bibfield  {journal} {\bibinfo  {journal} {Cancer Research}\ }\textbf
  {\bibinfo {volume} {44}},\ \bibinfo {pages} {4836s} (\bibinfo {year}
  {1984})}\BibitemShut {NoStop}%
\bibitem [{\citenamefont {Franken}\ \emph {et~al.}(2013)\citenamefont
  {Franken}, \citenamefont {Oei}, \citenamefont {Kok}, \citenamefont
  {Rodermond}, \citenamefont {Sminia}, \citenamefont {Crezee}, \citenamefont
  {Stalpers},\ and\ \citenamefont {Barendsen}}]{Franken}%
  \BibitemOpen
  \bibfield  {author} {\bibinfo {author} {\bibfnamefont {N.~A.}\ \bibnamefont
  {Franken}}, \bibinfo {author} {\bibfnamefont {A.~L.}\ \bibnamefont {Oei}},
  \bibinfo {author} {\bibfnamefont {H.~P.}\ \bibnamefont {Kok}}, \bibinfo
  {author} {\bibfnamefont {H.~M.}\ \bibnamefont {Rodermond}}, \bibinfo {author}
  {\bibfnamefont {P.}~\bibnamefont {Sminia}}, \bibinfo {author} {\bibfnamefont
  {J.}~\bibnamefont {Crezee}}, \bibinfo {author} {\bibfnamefont {L.~J.}\
  \bibnamefont {Stalpers}}, \ and\ \bibinfo {author} {\bibfnamefont {G.~W.}\
  \bibnamefont {Barendsen}},\ }\href {\doibase 10.3892/ijo.2013.1857}
  {\bibfield  {journal} {\bibinfo  {journal} {International Journal of
  Oncology}\ }\textbf {\bibinfo {volume} {42}},\ \bibinfo {pages} {1501}
  (\bibinfo {year} {2013})}\BibitemShut {NoStop}%
\bibitem [{\citenamefont {Dikomey}\ and\ \citenamefont {Jung}(1991)}]{Dikomey}%
  \BibitemOpen
  \bibfield  {author} {\bibinfo {author} {\bibfnamefont {E.}~\bibnamefont
  {Dikomey}}\ and\ \bibinfo {author} {\bibfnamefont {H.~W.}\ \bibnamefont
  {Jung}},\ }\href {\doibase 10.1080/09553009114550711} {\bibfield  {journal}
  {\bibinfo  {journal} {International journal of radiation biology}\ }\textbf
  {\bibinfo {volume} {59}},\ \bibinfo {pages} {815} (\bibinfo {year}
  {1991})}\BibitemShut {NoStop}%
\bibitem [{\citenamefont {Havemann}\ \emph {et~al.}(1987)\citenamefont
  {Havemann}, \citenamefont {Luinenburg}, \citenamefont {Wondergem},\ and\
  \citenamefont {Hart}}]{Havemann}%
  \BibitemOpen
  \bibfield  {author} {\bibinfo {author} {\bibfnamefont {J.}~\bibnamefont
  {Havemann}}, \bibinfo {author} {\bibfnamefont {M.}~\bibnamefont
  {Luinenburg}}, \bibinfo {author} {\bibfnamefont {J.}~\bibnamefont
  {Wondergem}}, \ and\ \bibinfo {author} {\bibfnamefont {A.}~\bibnamefont
  {Hart}},\ }\href {\doibase 10.1080/09553008714551031} {\bibfield  {journal}
  {\bibinfo  {journal} {International Journal of Radiation Biology and Related
  Studies in Physics, Chemistry and Medicine}\ }\textbf {\bibinfo {volume}
  {51}},\ \bibinfo {pages} {561} (\bibinfo {year} {1987})}\BibitemShut
  {NoStop}%
\bibitem [{\citenamefont {Kok}\ \emph {et~al.}(2014)\citenamefont {Kok},
  \citenamefont {Crezee}, \citenamefont {Franken}, \citenamefont {Stalpers},
  \citenamefont {Barendsen},\ and\ \citenamefont {Bel}}]{Kok}%
  \BibitemOpen
  \bibfield  {author} {\bibinfo {author} {\bibfnamefont {H.~P.}\ \bibnamefont
  {Kok}}, \bibinfo {author} {\bibfnamefont {J.}~\bibnamefont {Crezee}},
  \bibinfo {author} {\bibfnamefont {N.~A.}\ \bibnamefont {Franken}}, \bibinfo
  {author} {\bibfnamefont {L.~J.}\ \bibnamefont {Stalpers}}, \bibinfo {author}
  {\bibfnamefont {G.~W.}\ \bibnamefont {Barendsen}}, \ and\ \bibinfo {author}
  {\bibfnamefont {A.}~\bibnamefont {Bel}},\ }\href {\doibase
  https://doi.org/10.1016/j.ijrobp.2013.11.212} {\bibfield  {journal} {\bibinfo
   {journal} {International Journal of Radiation Oncology Biology Physics}\
  }\textbf {\bibinfo {volume} {88}},\ \bibinfo {pages} {739} (\bibinfo {year}
  {2014})}\BibitemShut {NoStop}%
\bibitem [{\citenamefont {Myerson}\ \emph {et~al.}(2004)\citenamefont
  {Myerson}, \citenamefont {Roti}, \citenamefont {Moros}, \citenamefont
  {Straube},\ and\ \citenamefont {Xu}}]{Myerson2004}%
  \BibitemOpen
  \bibfield  {author} {\bibinfo {author} {\bibfnamefont {R.~J.}\ \bibnamefont
  {Myerson}}, \bibinfo {author} {\bibfnamefont {J.~L.~R.}\ \bibnamefont
  {Roti}}, \bibinfo {author} {\bibfnamefont {E.~G.}\ \bibnamefont {Moros}},
  \bibinfo {author} {\bibfnamefont {W.~L.}\ \bibnamefont {Straube}}, \ and\
  \bibinfo {author} {\bibfnamefont {M.}~\bibnamefont {Xu}},\ }\href {\doibase
  10.1080/02656730310001609353} {\bibfield  {journal} {\bibinfo  {journal}
  {International Journal of Hyperthermia}\ }\textbf {\bibinfo {volume} {20}},\
  \bibinfo {pages} {201} (\bibinfo {year} {2004})}\BibitemShut {NoStop}%
\bibitem [{\citenamefont {Overgaard}(1984)}]{Overgaard6}%
  \BibitemOpen
  \bibfield  {author} {\bibinfo {author} {\bibfnamefont {J.}~\bibnamefont
  {Overgaard}},\ }\href {\doibase 10.3109/02841868409136001} {\bibfield
  {journal} {\bibinfo  {journal} {Acta Radiologica: Oncology}\ }\textbf
  {\bibinfo {volume} {23}},\ \bibinfo {pages} {135} (\bibinfo {year}
  {1984})}\BibitemShut {NoStop}%
\bibitem [{\citenamefont {Prabhu}\ and\ \citenamefont {Sharp}(2005)}]{Prabhu}%
  \BibitemOpen
  \bibfield  {author} {\bibinfo {author} {\bibfnamefont {N.~V.}\ \bibnamefont
  {Prabhu}}\ and\ \bibinfo {author} {\bibfnamefont {K.~A.}\ \bibnamefont
  {Sharp}},\ }\href {\doibase 10.1146/annurev.physchem.56.092503.141202}
  {\bibfield  {journal} {\bibinfo  {journal} {Annual Review of Physical
  Chemistry}\ }\textbf {\bibinfo {volume} {56}},\ \bibinfo {pages} {521}
  (\bibinfo {year} {2005})},\ \bibinfo {note} {pMID: 15796710}\BibitemShut
  {NoStop}%
\bibitem [{\citenamefont {Atkins}\ and\ \citenamefont
  {de~Paula}(2006)}]{Atkins}%
  \BibitemOpen
  \bibfield  {author} {\bibinfo {author} {\bibfnamefont {P.}~\bibnamefont
  {Atkins}}\ and\ \bibinfo {author} {\bibfnamefont {J.}~\bibnamefont
  {de~Paula}},\ }\href@noop {} {\emph {\bibinfo {title}
  {Atkins{\textquoteright} Pysical chemestry}}},\ \bibinfo {edition} {8th}\
  ed.\ (\bibinfo  {publisher} {W. H. Freeman and Company},\ \bibinfo {year}
  {2006})\BibitemShut {NoStop}%
\bibitem [{\citenamefont {Hillen}\ \emph {et~al.}(2010)\citenamefont {Hillen},
  \citenamefont {de~Vries}, \citenamefont {Gong},\ and\ \citenamefont
  {Finlay}}]{Hillen}%
  \BibitemOpen
  \bibfield  {author} {\bibinfo {author} {\bibfnamefont {T.}~\bibnamefont
  {Hillen}}, \bibinfo {author} {\bibfnamefont {G.}~\bibnamefont {de~Vries}},
  \bibinfo {author} {\bibfnamefont {J.}~\bibnamefont {Gong}}, \ and\ \bibinfo
  {author} {\bibfnamefont {C.}~\bibnamefont {Finlay}},\ }\href {\doibase
  10.3109/02841861003631487} {\bibfield  {journal} {\bibinfo  {journal} {Acta
  Oncologica}\ }\textbf {\bibinfo {volume} {49}},\ \bibinfo {pages} {1315}
  (\bibinfo {year} {2010})}\BibitemShut {NoStop}%
\bibitem [{\citenamefont {Naqa}\ \emph {et~al.}(2010)\citenamefont {Naqa},
  \citenamefont {Deasy}, \citenamefont {Mu}, \citenamefont {Huang},
  \citenamefont {J.}, \citenamefont {Lindsay}, \citenamefont {Apte},
  \citenamefont {Alaly},\ and\ \citenamefont {Bradley}}]{Naqa}%
  \BibitemOpen
  \bibfield  {author} {\bibinfo {author} {\bibfnamefont {I.~E.}\ \bibnamefont
  {Naqa}}, \bibinfo {author} {\bibfnamefont {J.~O.}\ \bibnamefont {Deasy}},
  \bibinfo {author} {\bibfnamefont {Y.}~\bibnamefont {Mu}}, \bibinfo {author}
  {\bibfnamefont {E.}~\bibnamefont {Huang}}, \bibinfo {author} {\bibfnamefont
  {H.}~\bibnamefont {J.}}, \bibinfo {author} {\bibfnamefont {P.~E.}\
  \bibnamefont {Lindsay}}, \bibinfo {author} {\bibfnamefont {A.}~\bibnamefont
  {Apte}}, \bibinfo {author} {\bibfnamefont {J.}~\bibnamefont {Alaly}}, \ and\
  \bibinfo {author} {\bibfnamefont {J.~D.}\ \bibnamefont {Bradley}},\ }\href
  {\doibase 10.3109/02841861003649224} {\bibfield  {journal} {\bibinfo
  {journal} {Acta Oncologica}\ }\textbf {\bibinfo {volume} {49}},\ \bibinfo
  {pages} {1363} (\bibinfo {year} {2010})}\BibitemShut {NoStop}%
\bibitem [{\citenamefont {Lea}\ and\ \citenamefont {Catcheside}(1942)}]{Lea}%
  \BibitemOpen
  \bibfield  {author} {\bibinfo {author} {\bibfnamefont {D.~E.}\ \bibnamefont
  {Lea}}\ and\ \bibinfo {author} {\bibfnamefont {D.~G.}\ \bibnamefont
  {Catcheside}},\ }\href {\doibase 10.1007/BF02982830} {\bibfield  {journal}
  {\bibinfo  {journal} {Journal of Genetics}\ }\textbf {\bibinfo {volume}
  {44}},\ \bibinfo {pages} {216} (\bibinfo {year} {1942})}\BibitemShut
  {NoStop}%
\bibitem [{\citenamefont {Eyring}(1935)}]{Eyring}%
  \BibitemOpen
  \bibfield  {author} {\bibinfo {author} {\bibfnamefont {H.}~\bibnamefont
  {Eyring}},\ }\href {\doibase 10.1063/1.1749604} {\bibfield  {journal}
  {\bibinfo  {journal} {The Journal of Chemical Physics}\ }\textbf {\bibinfo
  {volume} {3}},\ \bibinfo {pages} {107} (\bibinfo {year} {1935})}\BibitemShut
  {NoStop}%
\bibitem [{\citenamefont {Benzinger}(1983)}]{Benzinger}%
  \BibitemOpen
  \bibfield  {author} {\bibinfo {author} {\bibfnamefont {T.~H.}\ \bibnamefont
  {Benzinger}},\ }\href {\doibase 10.1152/ajpregu.1983.244.6.R743} {\bibfield
  {journal} {\bibinfo  {journal} {American Journal of Physiology-Regulatory,
  Integrative and Comparative Physiology}\ }\textbf {\bibinfo {volume} {244}},\
  \bibinfo {pages} {R743} (\bibinfo {year} {1983})},\ \bibinfo {note} {pMID:
  6859287}\BibitemShut {NoStop}%
\bibitem [{\citenamefont {Bischof}\ and\ \citenamefont {He}(2006)}]{Bischof}%
  \BibitemOpen
  \bibfield  {author} {\bibinfo {author} {\bibfnamefont {J.~C.}\ \bibnamefont
  {Bischof}}\ and\ \bibinfo {author} {\bibfnamefont {X.}~\bibnamefont {He}},\
  }\href {\doibase 10.1196/annals.1363.003} {\bibfield  {journal} {\bibinfo
  {journal} {Annals of the New York Academy of Sciences}\ }\textbf {\bibinfo
  {volume} {1066}},\ \bibinfo {pages} {12} (\bibinfo {year}
  {2006})}\BibitemShut {NoStop}%
\end{thebibliography}
\end{document}